%% file: main.tex
\documentclass[conference]{IEEEtran}

\usepackage{cite}
\usepackage{amsmath,amssymb,amsfonts}
\usepackage{algorithmic}
\usepackage{graphicx}
\usepackage{textcomp}
\usepackage{xcolor}

\usepackage{amsmath}
\usepackage{svg}
\usepackage{float}
\usepackage{color}
\usepackage{comment}
\usepackage{enumitem,xcolor}
\usepackage{algorithm}
\usepackage{algorithmic}
\usepackage{cuted} 

\newcommand{\nP}{\boxdot{\text{P}}} 
\newcommand{\nN}{\boxdot{\text{N}}} 
\newcommand{\nPPP}{\boxdot{\text{PPP}}} 
\newcommand{\nPPN}{\boxdot{\text{PPN}}} 
\newcommand{\nPNP}{\boxdot{\text{PNP}}} 
\newcommand{\nPNN}{\boxdot{\text{PNN}}} 
\newcommand{\nNPP}{\boxdot{\text{NPP}}} 
\newcommand{\nNPN}{\boxdot{\text{NPN}}} 
\newcommand{\nNNP}{\boxdot{\text{NNP}}} 
\newcommand{\nNNN}{\boxdot{\text{NNN}}} 

\newcommand{\kP}{\circledast{\text{P}}} 
\newcommand{\kN}{\circledast{\text{N}}} 
\newcommand{\kPPP}{\circledast{\text{PPP}}} 
\newcommand{\kPPN}{\circledast{\text{PPN}}} 
\newcommand{\kPNP}{\circledast{\text{PNP}}} 
\newcommand{\kPNN}{\circledast{\text{PNN}}} 
\newcommand{\kNPP}{\circledast{\text{NPP}}} 
\newcommand{\kNPN}{\circledast{\text{NPN}}} 
\newcommand{\kNNP}{\circledast{\text{NNP}}} 
\newcommand{\kNNN}{\circledast{\text{NNN}}} 

\usepackage{amsmath}
\usepackage{amssymb}
\usepackage{stmaryrd}
    
\begin{document}

\title{TA3: Testing Against Adversarial Attacks on Machine Learning Models}


\author{\IEEEauthorblockN{Yuanzhe Jin}
\IEEEauthorblockA{\textit{Department of Engineering Science} \\
\textit{University of Oxford}\\
Oxford, United Kingdom \\
yuanzhe.jin@eng.ox.ac.uk}
\and
\IEEEauthorblockN{Min Chen}
\IEEEauthorblockA{\textit{Department of Engineering Science} \\
\textit{University of Oxford}\\
Oxford, United Kingdom  \\
min.chen@oerc.ox.ac.uk}
}

\maketitle

\begin{abstract}
Adversarial attacks are major threats to the deployment of machine learning (ML) models in many applications. Testing ML models against such attacks is becoming an essential step for evaluating and improving ML models. In this paper, we report the design and development of an interactive system for aiding the workflow of Testing Against Adversarial Attacks (TA3). In particular, with TA3, human-in-the-loop (HITL) enables human-steered \emph{attack simulation} and visualization-assisted \emph{attack impact evaluation}. While the current version of TA3 focuses on testing decision tree models against adversarial attacks based on the One Pixel Attack Method, it demonstrates the importance of HITL in ML testing and the potential application of HITL to the ML testing workflows for other types of ML models and other types of adversarial attacks.    
\end{abstract}


\input{1_introduction}

\input{2_related_work}
\input{3_workflow}

\input{4_TA3_overview}

\input{5_TA3_techniques}

\input{6_case_studies}

\input{7_conclusion}

\bibliographystyle{IEEEtran}
\bibliography{ref}

\end{document}

%% file: 1_introduction.tex
\section{Introduction}
The rapid development of machine learning (ML) technology has started to bring forth the deployment of ML models in our daily lives, e.g., face recognition, traffic management, unmanned supermarkets, and autonomous vehicles. At the same time, the quality of ML models has received increasing attention. Research on \emph{adversarial attacks} has discovered a variety of vulnerabilities in ML models, ranging from
getting a model to mistake a panda for a gibbon \cite{Goodfellow:2014:Arx}
to getting Tesla's self-driving car to misjudge the speed limit sign \cite{Kim:2021:MLKE}.
In the field of ML, the current research effort on adversarial attacks places the main emphasis on discovering new adversarial attack methods and improving ML models' resistance against specific types of adversarial attacks. The former has resulted in a large collection of attacking methods \cite{Akhtar:2018:Access}, while the latter typically focuses on the training processes, e.g., data distill \cite{wang:2018:Arx} and adversarial training \cite{Bai:2021:IJCAI}.
In this work, we focus on ML testing processes against adversarial attacks.


There are differences as well as similarities between testing ML models and testing conventional software. The main differences include (i) deployable ML models are expected to make mistakes while deployable conventional software is expected to be bug-free; (ii) the inner workings of an ML model are expected to be difficult to comprehend, while a handcrafted program is expected to be fully understood. Because of these differences, the testing of ML models has relied primarily on statistical measures resulting from automated test runs, while the testing workflows for conventional software have relied extensively on human-in-the-loop (HITL).
Fundamentally, an ML model is also a program. Similar to a handcrafted program, the search space for a potential error in these programs is huge in that relying on a fully automated process to search for errors is mostly intractable computationally. Hence, in conventional software testing, HITL allows human experts to inject their knowledge to focus on highly probable parts of the search space. There is no reason why HITL cannot help human experts in testing ML models. This reasoning motivates us to develop an interactive software system to support ML testing workflows against adversarial attacks.  


There is a large collection of attacking methods, many of which apply to models trained with a specific algorithmic framework and specific types of training data \cite{Ren:2020:Eng}. There are also several strategies for defending against adversarial attacks, such as 
\emph{threat modeling}, \emph{attack simulation}, \emph{attack impact evaluation}, \emph{countermeasure design}, \emph{attack detection}, and \emph{defensive training} \cite{Chakraborty:2018:Arx}. These strategies may be implemented in individual techniques operating at different phases of ML workflows, e.g., for \emph{preparing data}, \emph{preparing learning}, \emph{training models}, and \emph{evaluating models} \cite{Sacha:2019:TVCG}.  
Hence, developing HITL-support techniques for supporting ML developers against adversarial attacks will be a long-term endeavor.
In this work, we focus on the phase of \emph{evaluating models} and the strategies of \emph{attack simulation} and \emph{attack impact evaluation}.
The software developed in this work currently features components for simulating adversarial attacks created by the One Pixel attack and components for visualizing decision-tree-based models.
With this work, we wish to make the following contributions:

\begin{itemize}
    \item We introduce a conceptual workflow for testing ML models against adversarial attacks, where through HITL-support techniques, ML experts can steer the simulation of potential attacks on an ML model and visualize the simulation results to evaluate the attack impact as well as the weak aspects of the model at the level of details that statistical measures cannot provide.
    \item We develop a prototype system, called TA3 (\emph{Testing Against Adversarial Attacks}), as an implementation of the conceptual workflow for testing decision tree models against One Pixel attacks, as well as for demonstrating the feasibility and the potential of the conceptual workflow.
    \item We propose a set of evaluation measures for quantifying the vulnerability of ML models, and demonstrate their uses as a statistical overview in conjunction with visual analysis.
\end{itemize}

%% file: 2_related_work.tex
\section{Related Work}
\subsection{Human-centered Artificial Intelligence}
Humans are increasingly exposed to artificial intelligence (AI) and machine learning (ML) techniques. Human-centered AI is an approach for making AI part of a larger system participated and managed by humans \cite{Auernhammer:2020:human}. It features a high level of human control, while computer automation improves human performance \cite{Riedl:2019:HBE}. In ML, although the machine may seem to do much of the work, there is a lot of human effort. Ramos et al. argued for more human-computer interactions (HCI) in ML processes \cite{Ramos:2019:ACM}. The establishment of the HCAI (Human-centered Artificial Intelligence) enables a pathway of AI toward human goals with more human participation in AI \cite{Shneiderman:2022:book}. On this basis, the concept of HCML (human-centered Machine Learning) has been developed \cite{Gillies:2016:ACM}.

In designing TA3, we drew inspiration from existing tools supporting HCAI and HCML. For instance, Tam et al. analyzed human contributions via interactive visualization in ML workflows \cite{Tam:2017:TVCG}, while McNutt and Chugh proposed pre-design templates for modular editing \cite{McNutt:2021:CHI}. Interfaces for ``fair'' ML were explored by Yan et al. and Ge et al., focusing on user-driven input modifications \cite{Yan:2020:ACM, Ge:2021:cast}. Zhang et al. contributed intelligent user interfaces for active learning and quality assurance in ML predictions \cite{Zhang:2021:TIIS, Zhang:2023:TIIS}. Additionally, Meng et al. developed VADAF for visualizing training dynamics \cite{Meng:2021:TIIS}, and Segura et al. supported collaborative sensemaking through visual narratives \cite{Segura:2021:TIIS}. Research by Chen et al. and Li et al. emphasized customizing interfaces and enhancing interactivity to improve user engagement \cite{Chen:2021:TIIS, Li:2023:TIIS}. Furthermore, studies by Yan et al. on user traits \cite{Yan:2023:TIIS} and Nakao on bias in interfaces \cite{Nakao:2022:TIIS} underscore the importance of personalization and inclusivity in interface design.
Additionally, we find studies that examine the correlation between decision-making style \cite{Hernandez:2023:TIIS} and user interface interactions, exploring how an individual's decision-making approach can impact their interaction preferences and experiences within a given interface.
Through this comprehensive exploration of user interface design and its intricate connections with human interactions, we aim to contribute to the advancement and optimization of user-centered applications, ultimately enriching user experiences.

Researchers made several attempts at the design process.
VINS \cite{Bunian:2021:ACM} takes UI images as input to retrieve visually similar design examples, allowing interface designers to better understand the needs of the design.
Several interactive tools have been developed to support different tasks, such as
 processing medical images \cite{Cai:2019:CHI},
 transferring text data into graph \cite{Thompson:2021:ACM},
 and time-series data \cite{Xu:2021:ACM}.
There are also user interface (UI) designs for supporting model developers and users, e.g., in
 natural language processing (NLP) \cite{Kang:2021:ACM},
 face recognition \cite{Wohler:2021:ACM},
 computer vision \cite{Balayn:2022:ACM,Zhang:2022:ACM}, and
 real-time games \cite{Dodge:2018:ACM}.
Some designs consider not only the tasks in ML workflows but also the people who use interactive visualization facilities (e.g.,  \cite{Hiniker:2019:ACM}). VIS4ML is an ontology that highlights various steps in ML workflows where visualization has been provided or can be provided to assist ML developers \cite{Sacha:2019:TVCG}.
Many researchers have proposed methods for visualizing various ML models \cite{Kahng:2018:TVCG, Ren:2017:TVCG, Sacha:2018:TVCG, Ming:2017:VAST, Hohman:2020:CHI, Krause:2017:VAST, Ye:2023:TVCG}. The most frequently mentioned ML models are deep neural networks \cite{Liu:2017:TVCG, Rauber:2017:TVCG, Pezzotti:2018:TVCG}. As one of the most widely used developing tools, TensorBoard in Tensorflow is an interactive tool \cite{Wongsuphasawat:2018:TVCG} for visualizing and editing neural network structures. Through interactive visualization, users can observe the data flowing running through the model.

Since many ML models \cite{LeCun:2015:Nature} are often considered difficult to understand, there are still many challenges for humans to apprehend the behavior and performance of ML models.
In this work, we focus on the family of decision tree models. 
In the field of visualization, researchers proposed many tree visualization methods.
For example, GoTree outlined a grammar \cite{Li:2020:CHI} for visualizing tree structures.
In response to the modification of decision trees, researchers propose a scalable decision tree visualization optimization method \cite{Elzen:2011:VAST}.
TreePOD \cite{Muhlbacher:2018:TVCG} and another research \cite{Teoh:2003:ACM} explored some of the decision tree generation processes that can be interactively managed by users.
For visualizing the training process, BOOSTVis uses the width of the edge to encode the amount of data passing through a decision path \cite{Liu:2018:TVCG}. BOOSTVis helps users analyze the training process visually to improve training efficiency.
We adopted their design of decision tree visualization in our work. RIT, the design of which was motivated by an ML application, provides a mathematically rigorous visual representation for depicting statistical values associated with tree nodes \cite{Jin:2024:TVCG}.

\subsection{Adversarial Attack Methods}
When ML models are deployed in applications, some small perturbations in input data that humans cannot see may have the ability to affect the accuracy of an ML model \cite{Zixiao:2021:WCMC}. This phenomenon was discovered by the Fast Gradient Signed Method (FGSM) proposed by Goodfellow et al. in 2014 \cite{Goodfellow:2014:Arx}. This method of perturbing input data indicates that the ML models are vulnerable to adversarial attacks.

Following their work, researchers proposed several derivative algorithms based on FGSM, such as
 RFGSM \cite{Tramer:2020:Arx},
 I-FGSM \cite{Kurakin:2016:Arx},
 DI2-FGSM \cite{Xie:2019:CVPR}, and so on,
to make FGSM more efficient in black-box attacks or improve the attack success rate in white-box attacks. For example, by adding randomness or momentum, FGSM can evade some model defenses.
In addition to FGSM, methods such as
 universal adversarial perturbations \cite{Moosavi:2017:CVPR},
 foolbox \cite{Rauber:2017:ARX},
 poison fog \cite{Shafahi:2018:CA}
can also interfere with the expected working results of ML models.
These algorithms are usually referred to as white-box attacks, as they introduce perturbation to input data based on the known model structures, parameters, or both.
There are also adversarial attack algorithms that can introduce perturbation without the knowledge of the ML models concerned (black box) \cite{Bhambri:2019:ARX}.

``One Pixel attack'' is a group of commonly studied adversarial attack algorithms. In each attack, such an algorithm selectively modifies a single pixel in an input image, often successfully causing an ML model to fail.
Su et al. proposed one such algorithm \cite{Su:2019:TEC}, which optimizes the selection of a pixel using differential evolution (DE) \cite{Storn:1997:GO}.
Kondratyuk provided open-source code for the algorithm \cite{Dan:Github}, which is used in this work.

\textcolor{black}{
Recent studies have highlighted various aspects of adversarial attacks in ML models. For instance, Garcia et al. \cite{Garcia:2023:SaTML} demonstrated that dimension reduction techniques can identify on-manifold adversarial examples. Hong et al. \cite{Hong:2023:SaTML} emphasized that while optimizing models for on-device efficiency, the increase in adversarial vulnerability cannot be overlooked. Apruzzese et al. \cite{Apruzzese:2023:SaTML} examined the gap between theoretical research and practical adversarial ML and pointed out that many real-world attackers do not rely on gradient computation, necessitating the need for more practical defense mechanisms. Croce et al. \cite{Croce:2024:SaTML} introduced prompt-agnostic adversarial attacks on segmentation models. Debenedetti et al. \cite{Debenedetti:2024:SaTML} presented techniques for evading black-box classifiers, demonstrating that effective attacks can be crafted without significant computational resources.}

\textcolor{black}{These findings underscore the importance of the evaluation and testing of ML models against adversarial attacks to ensure their robustness in both theoretical and practical applications.}

\subsection{Decision Tree}
As one of the widely used ML models, a decision tree is easy to understand \cite{Quinlan:1986:kap}. Users without much ML experience can appreciate different decision flows by observing nodes and edges in the decision tree. In ML testing, ML developers can evaluate a decision model by observing the data flow generated when a model processes the input data.

The earliest decision tree algorithm is Hunt’s Algorithm \cite{Breiman:1984:book}.
Built on the basic concept of Hunt’s Algorithm, several software systems were developed for training decision trees, including
  ID3 \cite{Quinlan:1986:kap},
  C4.5 \cite{Salzberg:ML:1994}, and
  CART \cite{Lewis:2000:book}.
Some methods such as GBDT \cite{Ye:2009:ACM}, LightGBM \cite{Ke:2017:NIPS}, and XGBoost tree \cite{Chen:2016:KDD} have been introduced to help decision tree building processes.
They use different gradient-boosting techniques to improve efficiency and accuracy in ML.
\textcolor{black}{In the context of secure and trustworth, Calzavar et al.~\cite{Calzavara:2023:SaTML} studied the problem of potential discrimination due to some sensitive features, and proposed an explainable approach for verifying global fairness of tree-based classifiers.}

In this work, we use the CART algorithm provided by scikit-learn \cite{Pedregosa:2011:ML} to train our ML models that are to be tested by using TA3. We assume that ML developers would test their models in a white-box manner. Training decision tree models using scikit-learn allows us to obtain the internal structures of the ML models, which can be visualized during the testing processes.







%% file: 3_workflow.tex
\section{A Conceptual Workflow}
\label{sec:Workflow}

\begin{figure*}[t]
  \centering
  \includegraphics[width=160mm]{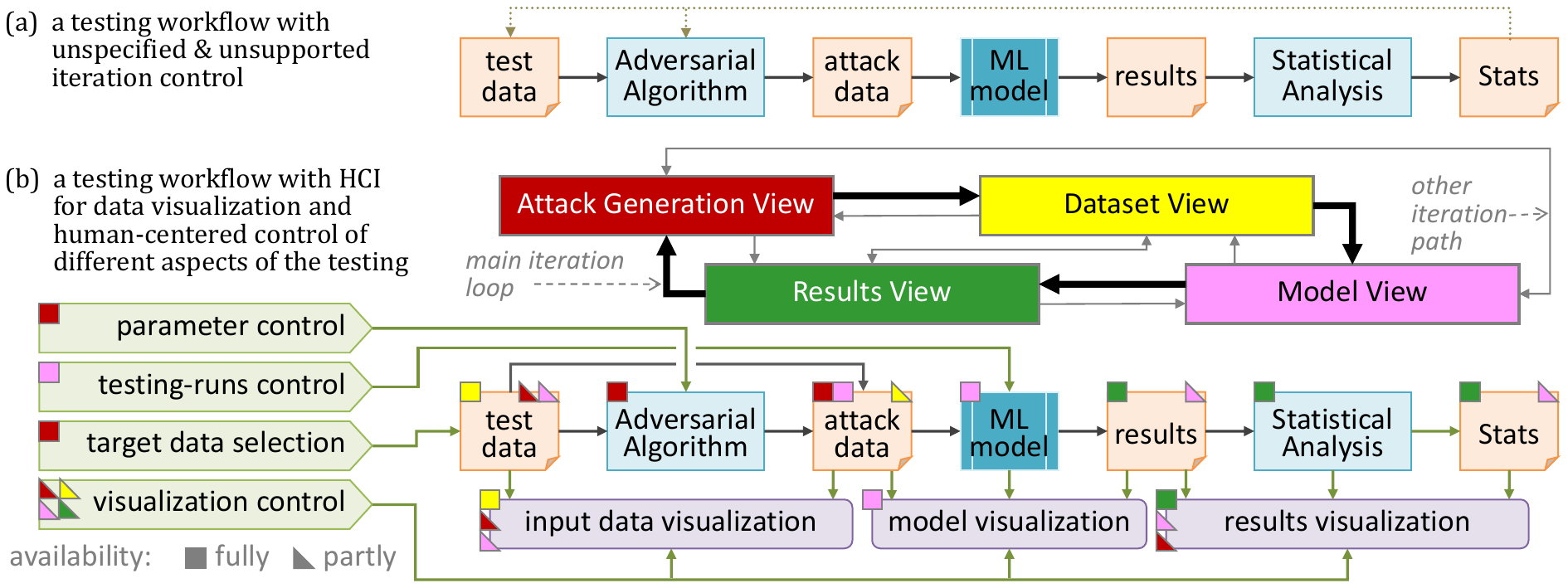}
  \caption{(a) Conventional workflows for testing adversarial attack algorithms focus on the statistics and instances that confirm the successes of an attack algorithm. Although iterative testing is common, such processes are typically not mentioned in the literature and are not supported by a purposely designed user interface.
  (b) An ideal workflow for testing against adversarial attacks should focus on the analysis of the behaviors of the model under attack and may feature many iterative loops where ML developers need to observe input data, generate attack data, carry out testing runs, and observe results. A software tool designed to support HITL activities can provide such iterative loops with effective and efficient user controls of the testing processes and visualization of the data, models, and results in the processes.}
  \label{fig:Workflows}
\end{figure*}

In the literature, almost all workflows for testing adversarial attack algorithms can be seen as an extension of the conventional ML testing workflows. As illustrated in Fig. \ref{fig:Workflows}(a), in such a workflow, selected testing data is perturbed by an adversarial attack algorithm, and the resultant attack data is then fed into an ML model. The ML developers typically evaluate the adversarial attack algorithm by observing some statistical measures about the attacks as well as some successful instances of the attack data. It is common for such a workflow to include many iterations, in which ML developers may select different test data to perturb or modify the parameters of the adversarial attack algorithm. Although the accumulated amount of human effort made in such iterative processes is non-trivial, it is usually not documented in the literature, and it is unsupported by any purposely built user interface to make the human effort more efficient and effective.

To a certain extent, ML developers have been accustomed to such iterative processes with limited software support for human-in-the-loop (HITL) activities. The objectives of testing adversarial attack algorithms are defined as (i) finding instances of effectual adversarial attacks, and (ii) obtaining the statistical measures for evaluating adversarial attack algorithms. However, if we focus the testing on evaluating the ability of a model to defend against adversarial attacks, we will need more than a few instances or a few statistical measures. Likely, the ML developers would like to obtain and observe information about the following questions:
\begin{itemize}
    \item[$Q_a$.] How easy for an ML model to be effectually attacked by an adversarial attack algorithm with some randomly selected parameters? 
    \item[$Q_b$.] How easy for an ML model to be effectually attacked by an adversarial attack algorithm with some randomly selected target data objects?
    \item[$Q_c$.] Are some types of target data objects more vulnerable than other types?
    \item[$Q_d$.] Are some parts of the ML model more vulnerable than other parts?
\end{itemize}

These requirements suggest the necessity of effective and efficient HITL support for several user tasks performed frequently in iterative testing processes, including:
\begin{enumerate}
    \item \textbf{Parameter Control} (cf. $Q_a$) --- with which an ML developer can assign some randomly selected parameters to an adversarial attack algorithm to observe the performance of the ML model under random attacks. The ML developer may also try to set parameters manually to see how easy to identify parameters that are ``harmfully'' effective for the attack algorithm.
    \item \textbf{Target Data Selection} (cf. $Q_{b,c}$) --- with which an ML developer can activate a random selection of target data objects to observe the performance of the ML model under random. The ML developer may also try to select target data objects manually to see how easy to identify some specific types of data objects as targets.
    \item \textbf{Testing-runs Control} (cf. $Q_{a,b,c,d}$) --- with which an ML developer can determine how many original data objects and how many perturbed data objects to be fed into the ML model.
    \item \textbf{Visualization Control} (cf. $Q_{a,b,c,d}$) --- with which an ML developer can interact with the visualization facilities (see below) to determine what data to be plotted, what levels of details are included and what feature to be highlighted, enabling exploratory observation and analysis.  
    \item \textbf{Input Data Visualization} (cf. $Q_{a,b,c}$) --- with which an ML developer can observe and analyze how a specific parameter setting affects the perturbation of the target data objects, what types of target data objects are selected, and which type may be more effectually attacked than other types.
    \item \textbf{Model Visualization} (cf. $Q_{c,d}$) --- with which an ML developer can observe and analyze how various parts of a model may behave differently in response to the original and perturbed data. If effectual attacks make certain parts of the model behave differently, these parts may be more vulnerable. The ML developer can also observe how different types of data affect different parts of the model differently. 
    \item \textbf{Results Visualization} (cf. $Q_{a,b,c,d}$) --- with which an ML developer can observe complex results from the testing runs of the model under different conditions. For example, ML developers may wish to observe the temporal patterns of effectual attacks among a sequence of attacks or compare multiple sequences of attacks by observing a variety of statistical measures associated with these sequences.
\end{enumerate}

Hence, we need to support the testing workflow in Fig. \ref{fig:Workflows}(a) by introducing a purposely built user interface, where these HITL-supporting facilities can be made available to ML developers as illustrated in the lower part of Fig. \ref{fig:Workflows}(b). Because there are many types of HITL-supporting facilities, some of which are shared by different tasks, we divide the testing workflow coarsely into four stages. As shown in the upper part of Fig. \ref{fig:Workflows}(b), the four stages are \emph{Attack Generation}, \emph{Data Observation}, \emph{Model Run \& Analysis}, and \emph{Results Analysis}.

Typically, when an ML developer starts a testing workflow, the first iteration will likely commence at the \emph{Data Observation} stage, and the ML developer may conduct a test with the original data objects without any perturbation (i.e., \emph{Model Run \& Analysis}). Having observed the ``normal'' behavior of the model (i.e., \emph{Results Analysis}), the ML developer may initiate a new iteration by generating some attack data at the \emph{Attack Generation} stage. The process will likely be repeated through multiple iterations with different algorithmic parameters and target data objects. Hence we use thick arrow-directed lines to indicate the main iteration loop. In practice, an ML developer may follow other paths as indicated by thin gray lines in the diagram. Therefore the switching among the four stages does not need to follow a particular order. This high-level design of the workflow for testing against adversarial attacks provides the design and development of TA3 with the basic conceptual framework as shown in Fig. \ref{fig:Workflows}(b).


%% file: 4_TA3_overview.tex
\section{TA3: System Overview}
\label{sec:TA3Overview}
While the conceptual workflow in Fig. \ref{fig:Workflows}(b) and the required HITL support for the workflow are generic to most processes for testing ML models against adversarial attacks, the design of some HITL-supporting facilities will need to be optimized for specific categories of ML models. The variations of the design may typically depend on (i) the type of input data to be processed by the ML model being tested since the data observation stage will require different visual representations for different types of data, (ii) the type of the ML model being tested since it may affect how the model is invoked and how the internal structure of the model is to be visualized, and (iii) the type of adversarial attack algorithm used in the testing since different types will likely have different sets of parameters. In this paper, we focus on a software system, TA3 (\emph{Testing Against Adversarial Attacks}), that demonstrates the feasibility of the conceptual workflow in Fig. \ref{fig:Workflows}(b), while supporting the testing processes against a commonly-studied family of adversarial attacks, i.e., pixel-based attacks on image classification models \cite{Chen:2017:WIFS, Yaratapalli:2021:ICTS}.

Pixel-based attacks introduce perturbation to imagery data by modifying one or more pixels in an image to be processed by an ML model. Since Szegedy et al. first reported such attacks in 2013 \cite{2013:ICLR}, many attacking algorithms have been reported in the literature. The TA3 incorporates the One Pixel attack algorithm \cite{Su:2019:TEC}, which uses the computation method of differential evolution method to find the pixel that maximizes the probability of an erroneous classification result. While the traditional testing workflow in Fig. \ref{fig:Workflows}(a) may be satisfied by finding the One Pixel, the testing against such attacks would like to find how an ML model would react to other similar attacks in addition to that One Pixel. Hence we adjust the One Pixel attack algorithm to generate multiple attacks and provide statistical and visual analysis of the performance of the ML model under such attacks.

\begin{figure*}[t]
  \centering
  \includegraphics[scale=0.31]{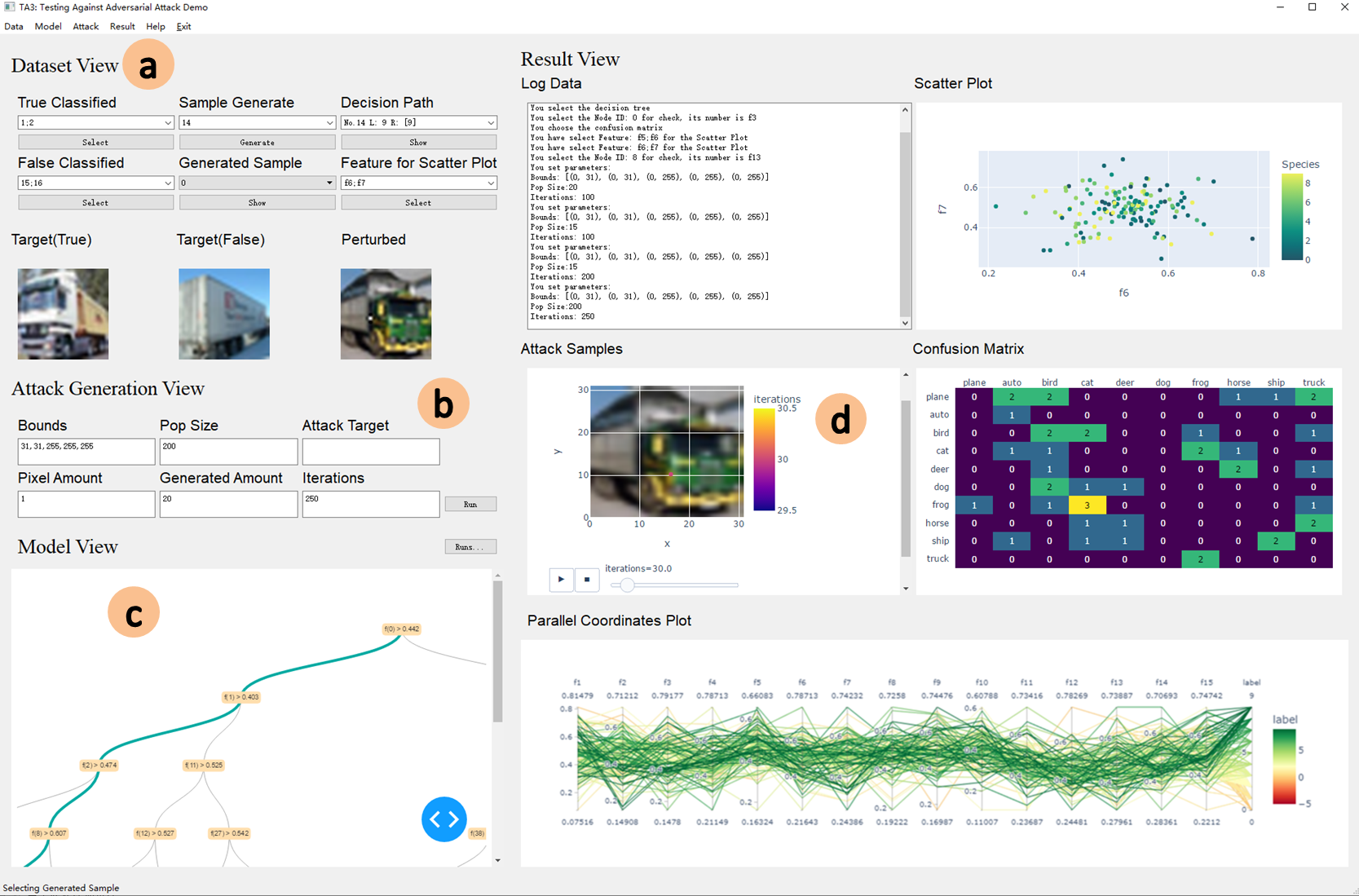}
  \caption{The main user interface of TA3, which is an HITL-supporting tool for testing ML models against adversarial attacks. The screen is roughly divided into four areas. (a) Below the menu bar, the \emph{Data View} area supports HITL activities related to the data used for testing ML models. (b) The \emph{Attack Generation View} area focuses on the HITL activities for controlling the adversarial attack algorithm and observing its input and output data. (c) The \emph{Model View} area enables the visualization of the testing data flowing through the model structure. (d) The \emph{Results View} area provides HITL-supporting facilities for observing and interacting with different visualization plots.
}
\label{fig:MainScreen}
\end{figure*}

TA3 was designed based on the conceptual workflow in Fig. \ref{fig:Workflows}(b). As shown in Fig. \ref{fig:MainScreen}, the main screen is organized into four parts, referred to as \emph{Dataset View} (top-left), \emph{Attack Generation View} (middle left), \emph{Model View} (bottom-left), and \emph{Results View} (right). These views host the most frequently-used HITL-supporting facilities and frequently-viewed information in the iterative testing processes. The users can access such facilities and information through the commands in the menu bar as well as a few short-cut command buttons in individual views. The four views correspond to the four stages in the upper part of Fig. \ref{fig:Workflows}(b). The \emph{Data View} was intentionally placed at the top-left part of the screen after some experimentation. The rationale of this design decision will be discussed below in the descriptions of \emph{Data View} and \emph{Attack Generation View}.

The \emph{Data View} provides HITL-supporting facilities for ($d_1$) loading a specific testing dataset, ($d_2$) viewing the summary information about the dataset, ($d_3$) selecting one or more target data objects in the dataset to be perturbed by the adversarial attack algorithm, ($d_4$) viewing the data objects in the dataset, ($d_5$) viewing the selected target data objects, ($d_6$), viewing the perturbed data objects, and ($d_7$) selecting the data objects (e.g., all or some data objects (original and/or perturbed)). The function ($d_1$) needs to be performed before performing any other tasks in the testing process, and partly because of this reason, \emph{Data View} was placed at the top-left part of the screen, which is naturally a place to start HITL actions.

\begin{figure*}[t]
  \centering
  \begin{tabular}{@{}c@{\hspace{8mm}}c@{}}
       \includegraphics[height=41mm]{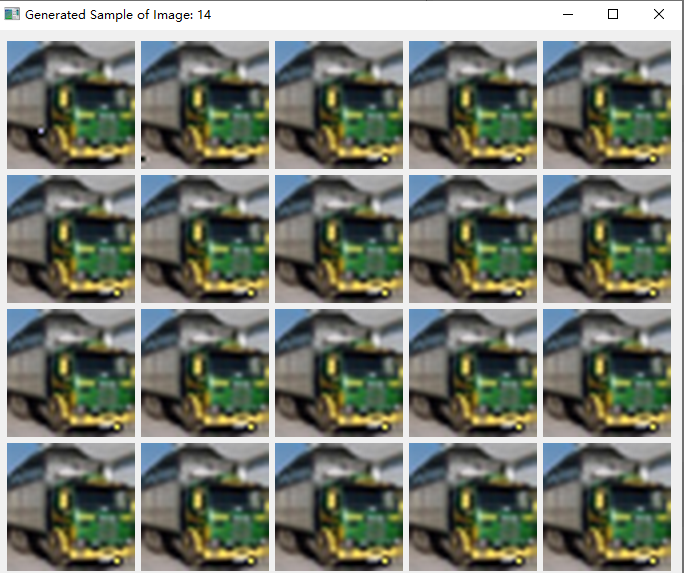} & 
       \includegraphics[height=41mm]{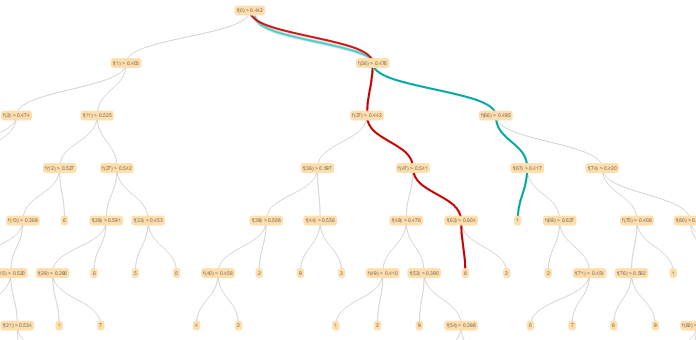}
  \end{tabular}
  \caption{A pop-up window showing all perturbed data objects generated by the adversarial attack algorithm. Left: A pop-up window showing all perturbed data objects. Right: A decision tree with true- and false-positive data flows.}
  \label{fig:TwoExamples}
\end{figure*}

\emph{Attack Generation View} provides HITL-supporting facilities for ($g_1$) setting and modifying the parameters of the adversarial algorithm, ($g_2$) viewing the input data to the algorithm (i.e., the target data), and ($g_3$) viewing the perturbed data (e.g., the results of applying the algorithm to the target data). The functions ($g_2$) and ($g_3$) are more or less the same as ($d_4$) and ($d_5$), and therefore, the visual representations of these data objects are shared between the \emph{Data View} and the \emph{Attack Generation View}. As the adversarial attack algorithm usually generates many perturbed samples from the same target data object, TA3 provides three visual representations for displaying the perturbed data, including (i) a static representation of a perturbed data object that was generated first by the adversarial attack algorithm, (ii) an animation of all perturbed data objects, and (iii) a static grid view of all perturbed data objects. The first two representations are available on the main screen, while the third representation is available on demand through a pop-up window as shown in Fig.~\ref{fig:TwoExamples}(left).

\emph{Model View} provides HITL-supporting facilities for ($m_1$) controlling the execution runs of the ML model being tested and ($m_2$) investigating how the model predictions are affected by the input data objects and the internal structure of the model with the aid of model visualization. The design for ($m_2$) depends on two main factors, namely whether the ML model is available to the testing processes as a black-box model or while-box model, and if it is a white-box model, which model type (e.g., decision tree, CNN, etc.). The TA3 system was designed for ML developers to evaluate their models in the decision tree family. We thus consider only white-box decision-tree models. The internal structure of the model and its data flow from input to output can be observed using tree visualization. For example, as illustrated in Fig.~\ref{fig:TwoExamples}(right), users can visualize the data flow for true- and false-positives to see how adversarial attacks redirect some part of the true-positive data flow into wrong paths.

\emph{Results View} provides HITL-supporting facilities for ($r_1$) viewing the testing results and summary statistics through a variety of visualizations. For example, as shown in Fig. \ref{fig:MainScreen}, there is a scatter plot for displaying the features used in the decision tree, a matrix plot for displaying a confusion matrix derived from the testing results, a parallel coordinates plot for showing the testing data objects and their feature values. In Section \ref{sec:TA3techniques}, we will further discuss the visualization plots and summary statistics available in TA3 in detail.

%% file: 5_TA3_techniques.tex
\section{TA3: Algorithm, Interaction, Statistics, and Visualization}
\label{sec:TA3techniques}
Through the user interface as shown in Fig. \ref{fig:MainScreen}, TA3 provides ML developers with access to algorithms, interaction (interactive mechanisms), statistics, and visualization. In this section, we detail these four technical aspects of TA3. 

\subsection{Algorithms: Machine Learning and Attack Simulation}
In TA3 and indeed any software of this nature, there are many algorithms. Here we focus on two families of algorithms, namely machine learning (ML) and attack simulation. Because this work relies on such algorithms available in the public domain, we describe them briefly here. 

In the context of ML, there are two main types of algorithms. A trained ML model is an algorithm that is partly designed by human developers (e.g., the structure of a neural network or a set of features of a decision tree) and partly learned from the training data.
Meanwhile, an ML software platform for training an ML model hosts the core algorithm that embodies an ML technique, which may be unblended or hybrid and may also be referred to as an ML method or an ML framework.

In this work, we focus on decision tree models, which are particularly suitable for white-box testing. We trained our ML models using the CART algorithm in the scikit-learn Library \cite{Pedregosa:2011:ML}. Given a set of pre-processed feature data, CART constructs a binary tree iteratively by selecting the most information-rich feature in the iteration concerned and determining the optimal threshold for splitting the data flow into two branches.

For white-box testing, TA3 displays the decision tree structure as well as the relationships between a list of features and data objects (i.e., images in this case) in a parallel coordinates plot. TA3 can accommodate any number or type of features. As an example, the models shown in this paper are trained with feature data obtained by using principle component analysis (PCA) \cite{Maćkiewicz:1993:CG, Bodine:2022:SNCS}. For each image in a training dataset, we use fifteen principle components as features for training each model. These models can easily be replaced by decision tree models that are trained with other types of features such as HOG \cite{Dalal:2005:CVPR} or SIFT \cite{Lowe2004} features. Users can also apply their methods.
The datasets used for ML in this work are Cifar-10 \cite{krizhevsky:2009:web}, Fashion-MNIST \cite{xiao:2017:Arx}, MNIST \cite{Lecun:1998:IEEE}, while TA3 currently can handle any decision tree model for image classification and any image dataset for testing such a model.%

Algorithms for adversarial attacks are mostly hand-crafted by experts though most algorithms contain optimization processes that may be regarded as a kind of ML. There are many adversarial attack algorithms in the literature and most algorithms have user-adjustable parameters. Different algorithms and different parameters result in different performances, making the testing processes important as well as challenging.

In this work, we focus on one particular attacking algorithm, i.e., the One Pixel attack. 
We used the open-source implementation provided by Kondratyuk (from Google Research) who made the code available at GitHub \cite{Dan:Github}. The implementation was based on the original algorithm proposed by Su et al. \cite{Su:2019:TEC}. Because it is an open-source implementation, we can easily make a small modification to extract some interim data for visualization and to generate multiple attacks for testing ML models.

\subsection{Interaction: Testing Steering}
As shown in Fig. \ref{fig:Workflows}(b), the HITL approach is essential for target data selection, parameter control (for the adversarial attack algorithm), testing runs control, and visualization control. In general, human-computer interaction (HCI) is an elementary component of HITL. Without HCI, ML developers will not be able to steer the testing process based on their knowledge of the ML models, datasets, and adversarial attack algorithms that are being tested, and their observation and interpretation of the statistics and visualization. 

Some of the widgets for these HCI tasks are accessible on the main screen shown in Fig. \ref{fig:MainScreen}, while others are accessible in the corresponding pop-up windows. These HCI tasks are:
\begin{itemize}
    \item[] \hspace{-3.5mm}\textbf{Target Data Selection}
    \item Dataset Selection -- for selecting a dataset to be loaded into the system.
    \item Candidate Data Object Selection -- for selecting two original data objects (images in this work), one is correctly classified by the ML model, and another is incorrectly classified by the model. The correctly classified data object is usually the target of an adversarial attack algorithm. The incorrectly classified data object is a source of risk, perhaps with a more serious one since it is likely easier to perturb.
    \item Target Data Object Selection -- for selecting an image to be perturbed.
    \item[] \hspace{-3.5mm}\textbf{Parameter Control}
    \item Bound -- for restricting the 5D search space for a pixel in an RGB image, i.e., $x$, $y$, $r$, $g$, $b$.
    \item Pop Size -- for defining the population size of the differential evolution method.
    \item Target Data Object Selection -- for selecting an image to be perturbed.
    \item Attack Target -- for selecting the target to be perturbed. If it is left blank, it is an untargeted attack.
    \item Num. Pixels -- for controlling the number of pixels to be perturbed per attack.
    \item Num. Attacks -- for controlling the number of adversarial images to be generated.
    \item Iterations -- for controlling the number of iterations for finding the pixels to be perturbed.
    \item[] \hspace{-3.5mm}\textbf{Testing Runs Control}
    \item Data Flow Selection -- for selecting the data flow of a set of images to be tested, with options for various combinations of the target images (0, 1, or 2), generated attack images (0 $\sim$ all), and a subset of the loaded testing data (0 $\sim$ all). This also defines what is to be visualized in the decision tree visualization.
    \item Run -- for activating a simple testing run based on the Data Flow Selection 
    \item Runs ... -- for specifying more complex testing involving multiple target images.
    \item[] \hspace{-3.5mm}\textbf{Visualization Control}
    \item Features for the scatter plot -- for selecting two features to be visualized in the scatter plot.
    \item Axes movement -- for moving the axes within the parallel coordinate plot to enable the easy observation of feature relations.
    \item Brushing -- for selecting and highlighting a subset of data objects (i.e., lines). 
\end{itemize}

Although visualization control seems to be for viewing the testing results rather than steering the testing, visualizing the testing results in one testing iteration will influence the users' decisions about the next iteration and subsequent iterations. Hence visualization control is a form of indirect testing steering.

\begin{table*}[t]
    \centering
    \caption{Six new measures designed for testing a model's ability to deal with adversarial attacks.}
    \label{tab:NewMeasures}
    \vspace{-6mm}
\normalsize
\begin{align*}
    \textbf{Measure}\quad
        & \textbf{General}
        &&\textbf{Positive-specific}
        &&\textbf{Negative-specific}\\
    \textbf{model-robustness rate:}\quad
        &\frac{\nPPP + \nPNN + \nNNN + \nNPP}{\nP + \nN}
        &&\frac{\nPPP + \nPNN}{\nP}
        &&\frac{\nNNN + \nNPP}{\nN}\\[2mm]
    \textbf{attack-breach rate:}\quad
        &\frac{\nPPN + \nPNP + \nNNP + \nNPN}{\nP + \nN}
        &&\frac{\nPPN + \nPNP}{\nP}
        &&\frac{\nNNP + \nNPN}{\nN}\\[2mm]
    \textbf{adversarial-impact rate:}\quad
        &\frac{\nPPN + \nNNP}{\nP + \nN}
        &&\frac{\nPPN}{\nP}
        &&\frac{\nNNP}{\nN}\\[2mm]
    \textbf{attack-failure rate:}\quad
        &\frac{(\nP - \nPPN)  + (\nN - \nNNP)}{\nP + \nN}
        &&\frac{\nP - \nPPN}{\nP}
        &&\frac{\nN - \nNNP}{\nN}\\[2mm]
    \textbf{intended-perturbation rate:}\quad
        &\frac{\nPPN + \nNNP}{\nPPN + \nPNP + \nNNP + \nNPN}
        &&\frac{\nPPN}{\nPPN + \nPNP}
        &&\frac{\nNNP}{\nNNP + \nNPN}\\[2mm]   
    \textbf{unintended-perturbation rate:}\quad
        &\frac{\nPNP + \nNPN}{\nPPN + \nPNP + \nNNP + \nNPN}
        &&\frac{\nPNP}{\nPPN + \nPNP}
        &&\frac{\nNPN}{\nNNP + \nNPN}\\[2mm]
\end{align*}
\vspace{-8mm}
\end{table*}

\subsection{Statistics: Simulation Summarization}
\label{sec:Statistics}
Research papers on adversarial attacks typically evaluate an attack algorithm by measuring the statistics about the successful attacks (e.g., \cite{Chen:2022:TPAMI}) and the difference between the accuracy of original testing data and the attacked testing data (e.g., \cite{Moosavi-Dezfooli_2016_CVPR}).
Although such statistical measures support the basic evaluation, they are much too coarse in comparison with commonly used statistical measures in computer vision (e.g., precision, recall, F1-score, and so on).
In this work, we define several new statistical measures to enable ML developers to conduct a finer evaluation.

Consider a binary classification model $\mathbf{M}$ and a set of testing data objects $o_1, o_2, \ldots, o_n$, each of which has a ground truth label ``P'' (positive) or ``N'' (negative). When the model $\mathbf{M}$ is tested against an original data object, $o_i$, it may return ``P'' or ``N''. Similarly, when $\mathbf{M}$ is tested against an attacked data object $\alpha(o_i)$, it also returns either ``P'' or ``N''. Hence, for each of the $n$ data objects, its testing results can be represented by three letters, representing $\langle\text{ground truth}\rangle\langle\text{test against}\;o_i\rangle\langle\text{test against}\;\alpha(o_i)\rangle$ respectively. For example, ``PPN'' indicates that a data object has a ground truth label ``P'', $\mathbf{M}$ classifies it correct as ``P'' (true positive), and $\mathbf{M}$ classifies the attacked object as ``N'' (false negative).
There are eight possible cases, ``PPP'', ``PPN'', ``PNP'', ``PNN'', ``NPP'', ``NPN'', ``NNP'', and ``NNN''.

Let us add a symbol ``$\boxdot$'' or ``$\boxdot_n$'' in front of these to denote the number of each case in testing a total of $n$ data objects. Let us first define two simplified notations:
\[\nP = \nPPP + \nPPN + \nPNP + \nPNN\] 
\[ \nN = \nNPP + \nNPN + \nNNP + \nNNN \]
We can easily write down some commonly-used statistical measures for testing $\mathbf{M}$ against the original data and the attacked data as:
%

\noindent$\blacksquare$ \textbf{Original Data Statistics}
\begin{align*}
    \textbf{accuracy:}\qquad
        & \frac{\nPPP + \nPPN + \nNNP + \nNNN}{\nP + \nN} \\[1mm]
    \textbf{precision:}\qquad
        & \frac{\nPPP + \nPPN}{\nPPP + \nPPN + \nNPP + \nNPN} \\[1mm]
    \textbf{recall:}\qquad
        & \frac{\nPPP + \nPPN}{\nP} \\[1mm]
    \textbf{F1 score:}\qquad
        & \frac{2\cdot\nPPP + 2\cdot\nPPN}{\nP + \nPPP + \nPPN + \nNPP + \nNPN}
\end{align*}

\noindent$\blacksquare$ \textbf{Attacking Data Statistics}
\begin{align*}
        \textbf{accuracy:}\qquad
        & \frac{\nPPP + \nPNP + \nNPN + \nNNN}{\nP + \nN}\\[1mm]
    \textbf{precision:}\qquad
        & \frac{\nPPP + \nPNP}{\nPPP + \nPNP + \nNPP + \nNNP}\\[1mm]
    \textbf{recall:}\qquad
        & \frac{\nPPP + \nPNP}{\nP}\\[1mm]
    \textbf{F1 score:}\qquad
        & \frac{2\cdot\nPPP + 2\cdot\nPNP}{\nP + \nPPP + \nPNP + \nNPP + \nNNP}\\
\end{align*}

Essentially, the original statistics ignore the last letter by considering, e.g., true positives as $\nPPP + \nPPN$, since the last letter is about the predictions $\mathbf{M}$ for the attacked objects. Meanwhile, the attacking statistics ignore the middle letter (e.g., true positives $= \nPPP + \nPNP$), since the middle letter is about the predictions $\mathbf{M}$ for the original objects. 

In addition, we can introduce new statistical measures to summarize the successful and failed attacks as well as intended and unintended perturbations in the testing results. The formulae of these new measures are given in Table \ref{tab:NewMeasures}.

The measure \textbf{model-robustness rate} summarizes the ability of $\mathbf{M}$ to resist the adversarial attack algorithm by considering those predictions of $\mathbf{M}$ that remain the same before and after the attacks.
The measure \textbf{attack-breach rate} summarizes the level of success of the adversarial attack algorithm in altering the predictions of $\mathbf{M}$.
We have \textbf{attack-breach rate} $= 1 -$ \textbf{model-robustness rate}.
Among such perturbations, some cause $\mathbf{M}$ to change correct decisions to incorrect decisions, and others vice versa.
The measure \textbf{adversarial-impact rate} summarizes the actual adversarial damages by excluding those attacks that have unintentionally corrected the errors made by a model when applying it to the unperturbed data, i.e. $\nPNP$ and $\nNPN$.
Meanwhile, the measure \textbf{attack-failure rate} summarizes the failed attacks, such that \textbf{attack-failure rate} $= 1 -$ \textbf{adversarial-impact rate}.
The measure \textbf{intended-perturbation rate} summarizes the damaging attacks among the successful attacks, while \textbf{unintended-perturbation rate} summarizes those successful attacks that unintentionally cause $\mathbf{M}$ to change incorrect decisions to correct decisions. We have \textbf{unintended-perturbation rate} $= 1 -$ \textbf{intended-perturbation rate}.

The above measures are all defined in the same context of commonly-used statistical measures in machine learning by assuming each data object $o_i$ is attacked only once (cf. only tested once). As discussed in previous subsections, in practice, adversarial attacks will likely be applied to the same data object multiple times, and TA3 facilitates the testing against such multi-attacks. We can define a similar set of single-object and multi-attack measures by adapting the above multi-object and single-attack measures. 

Let us first consider a single data object $o$ being attacked multiple times, different attacks attenuate the data object differently. When $o$ is attacked for $k$ times, there will be $k$ modified data objects. In many ways, this is similar to having a testing dataset that contains $k$ copies of the same data objects. To distinguish the difference between testing $n$ different data objects with a single attack on each of them and testing the same data object with $k$ attacks, we introduce a new symbol $\circledast$ (instead of $\boxdot$) to indicate the number qualities in the statistical measures. For example, consider an object with the ground truth label ``P'' and it would be correctly classified as ``P'' by $\mathbf{M}$ normally.
$\kPPN$ indicates among $k$ attacks, the number of the attacks that made $\mathbf{M}$ returning the label ``N''.

It is necessary to note that the measures for single-object and multi-attacks depend only on the number of quantities related to the specific ground truth label of the data object. For example, if the ground truth label is ``P'', then $\kP = k$ and $\kN = \kNPP = \kNPN = \kNNP = \kNNN = 0$. The negative-specific measures do not apply to the test of this data object. If the $\mathbf{M}$ could classify the original object correctly and $s\,(s \leq k)$ attacks were successful, we would have $\kPPN = s$, $\kPPP = k-s$, and $\kPNP = \kPNN = 0$.
We have \textbf{attack-breach rate}$(\circledast_k) =$ \textbf{adversarial-impact rate}$(\circledast_k) = s/k$, and
the \textbf{model-robustness rate}$(\circledast_k) =$ \textbf{attack-failure rate}$(\circledast_k) = 1-s/k$.

Note that the square in $\boxdot$ symbolizes a dataset, while the circle in $\circledast$ symbolizes a data object. The dot indicates a single attack per object, while the $\ast$ indicates multiple attacks. By extrapolation, we can also use $\boxast$ to symbolize statistical measures obtained by testing $n$ data objects, each of which receives $k$ attacks.
In many situations, the computer screen may display a measure without a detailed formula. Therefore, we always indicate the context of the test, with at least one of the three symbols $\boxdot$, $\circledast$, and $\boxast$. If there is sufficient display space, we will indicate the numbers of objects and attacks, e.g., \textbf{attack-impact rate}($\boxast_{1000, 10}$) = 0.52 or \textbf{attack-impact rate}(1000 objects, 10 attacks) = 0.52.

Although these new statistical measures offer more information than the traditional measures such as success rate, model developers may wish to observe some details about multi-attacks. Visualization can deliver more information to complement statistics that may abstract data too quickly \cite{Chen:2011:C}. This will be discussed in the next subsection.   

\subsection{Visualization: Simulation Progression and Testing Results}
\label{sec:Visualization}
As shown in Fig. \ref{fig:MainScreen}, visualization plots enable ML developers to observe a variety of visualization plots, which display the data captured during model testing as well as the summary statistics about the model performance. Broadly, the data that can be visualized in TA3 falls into the following categories:

\begin{figure}[t]
    \centering
    \begin{tabular}{@{}c@{\hspace{1mm}}c@{\hspace{2mm}}c@{}}
    \includegraphics[height=22mm]{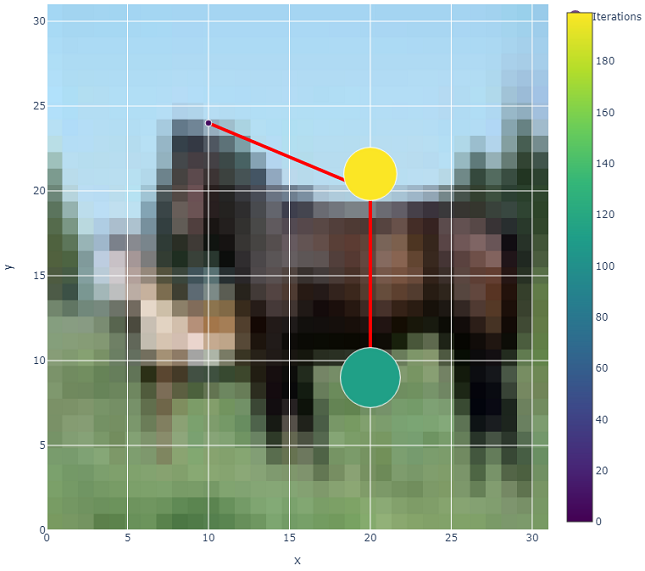} &
    \includegraphics[height=22mm]{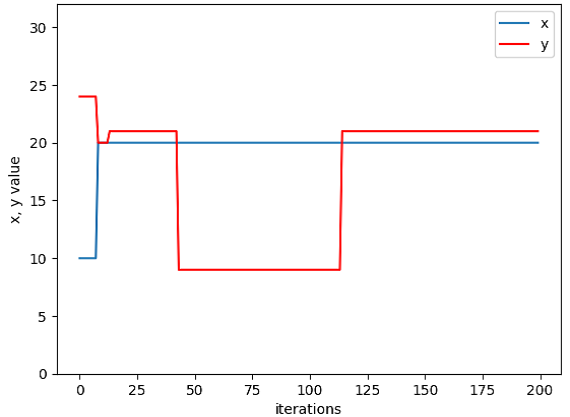} &
    \includegraphics[height=22mm]{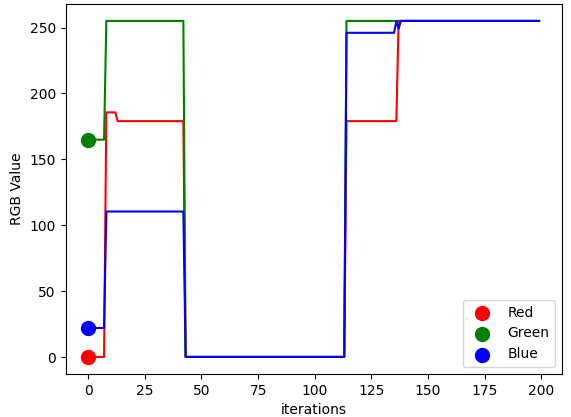} \\
    \small (a) path on image &
    \small (b) $x, y$-coordinate &
    \small (c) RGB values
  \end{tabular}
    \caption{Two ways to visualize the movement of the simulated attack points during a test. (a) Intuitive view in the context of the image (Image No. 75) being attacked. (b \& c) Detailed progression of the attacking points during the simulation, with color-mapped pixels representing the amount of increasing and decreasing pixel brightness and a yellow background indicating a successful attack.}
    \label{fig:P-Attacks}
\end{figure}

\begin{figure}[t]
  \centering
  \begin{tabular}{@{}c@{\hspace{2mm}}c@{\hspace{2mm}}c@{}}
    \includegraphics[width=28mm,height=15mm]{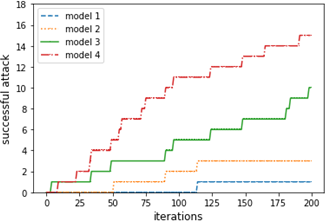} &
    \includegraphics[width=28mm, height=15mm]{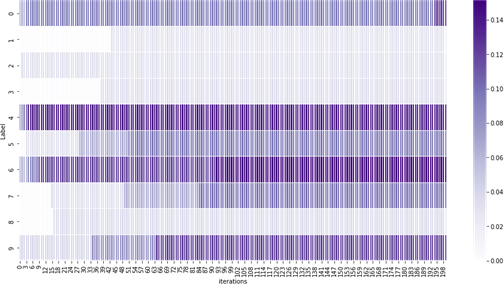} &
    \includegraphics[width=28mm,height=15mm]{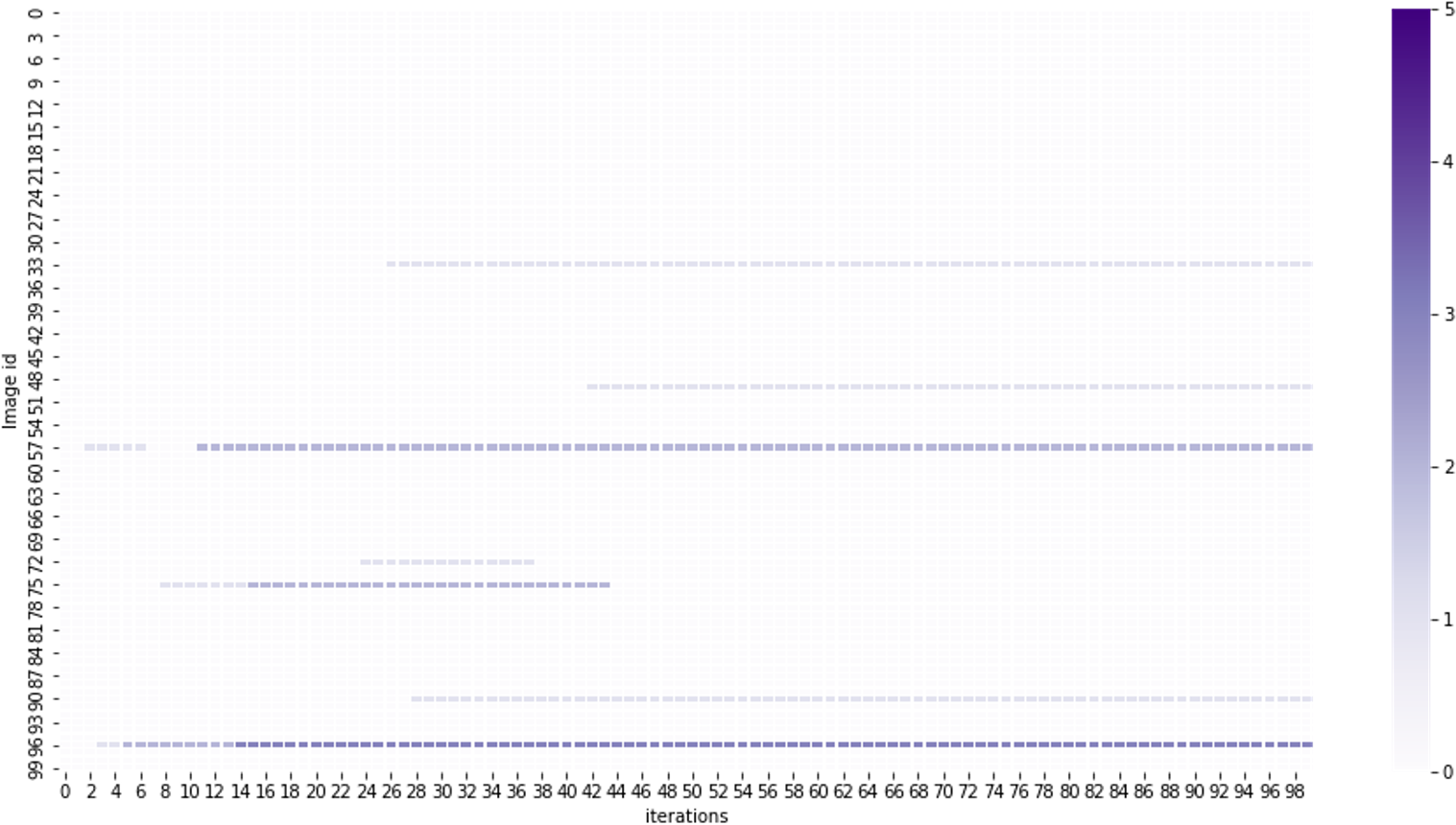}\\
    \small (a) four models & \small (b) 10 classes & \small (c) 100 images
  \end{tabular}
  \caption{Three visualization plots for comparing (a) four models in terms of the number of cumulative successes vs. iterations, (b) compare 10 classes in terms of success rate vs. the number of iterations, and (c) compare 100 images in terms of success rate vs. the number of iterations. }
  \label{fig:P-Statistics}
\end{figure}

\begin{figure}[t]
    \centering
    \begin{tabular}{@{}c@{\hspace{3mm}}c@{}}
    \includegraphics[width=33mm,height=20mm]{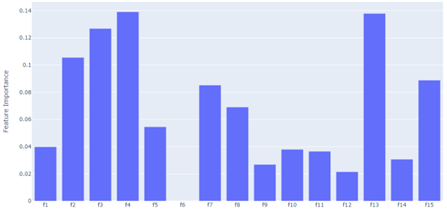} &
    \includegraphics[width=46mm,height=20mm]{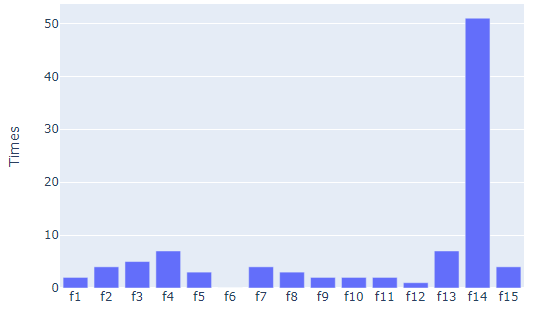}\\
    \small (a) feature importance &
    \small (b) feature usage in nodes 
  \end{tabular}
    \caption{Two example visualizations of model statistics.}
    \label{fig:ModelStatistics}
\end{figure}

\begin{itemize}
    \item \textbf{Simulation progression: attacking patterns} --- During simulation, the perturbed data objects can be visualized directly as shown on the left of Fig. \ref{fig:P-Attacks}. The movement of the attacking pixels can also be visualized as a path superimposed on an enlarged image and as two time series for $x$- and $y$- coordinates. The former enables the observation of attacking pixels in conjunction with the imagery context, while the latter enables the observation of whether there are repeated attacks in a small area.
    \item \textbf{Simulation progression: evolving statistics} --- During simulation, various statistical measures are computed for each iteration. As exemplified in Fig. \ref{fig:P-Statistics}, we can observe (a) the number of successful attacks on the same image during testing different models, (b) the vulnerability of images in different classes, and the vulnerability of individual images. We will show more examples of such plots deployed in our case studies in Section \ref{sec:CaseStudies}.
    \item \textbf{Testing results: model statistics} --- after simulation, various statistical measures are made available through visualization plots as well as numerical values. This includes some commonly-used ML statistics such as accuracy, precision, recall, F1-scores, and confusion matrix as well as the statistical measures introduced in Section \ref{sec:Statistics} for summarizing adversarial attacks. For example, a confusion matrix is shown in the middle-right of Fig. \ref{fig:MainScreen}, while two bar charts are shown in Fig. \ref{fig:ModelStatistics}, one for summarizing the importance of features used by an ML model and another to show feature used in the nodes.
    \item \textbf{Testing results: analytical visualization} --- Some visualization plots are designed to support detailed analysis of ML models being tested. As shown in Fig. \ref{fig:MainScreen}, there are scatter plots for observing the possible correlational or independent relations between two features (top right), parallel coordinates plots for examining such relations across many features (bottom right), and data flows in a decision tree (bottom-left).   
\end{itemize}

The types of plots currently available in TA3 include:
\begin{itemize}
    \item bar chart --- for visualizing the overall statistics for different categories of testing data. Fig. \ref{fig:ModelStatistics} shows two such examples. Other uses of this form of visualization include average success rates for different classes, models, datasets, and so on.
    \item line plot --- for visualizing the evolution of various statistical measures during simulation, such as the number of successful and/or unsuccessful attackers ($y$-axis) at each iteration as shown in Fig. \ref{fig:P-Statistics}(a). Multiple lines can be used to represent different conditions controlled by categorical variables, such as multiple models, multiple datasets, and classes, as well as a small number of discrete instances of a numerical control parameter.
    \item scatter plot --- for visualizing the relationships between two features (e.g., top-right of Fig. \ref{fig:MainScreen} and (b) in Fig. \ref{fig:P-Attacks}).
    \item pixel-based plot --- for visualizing the confusion matrix of the testing results (e.g., middle-right of Fig. \ref{fig:MainScreen}), numerical indicators about the testing results in terms of different categories of data objects (e.g., (b) and (c) in Fig. \ref{fig:P-Statistics}).
    \item node-link tree plot --- for visualizing the testing data that flows through a decision tree model (e.g., bottom-left of Fig. \ref{fig:MainScreen}).
    \item parallel coordinates plot --- for visualizing the testing data objects and features used by the decision models (e.g., bottom-left of Fig. \ref{fig:MainScreen}).
\end{itemize}

%% file: 6_case_studies.tex
\section{Case Studies}
\label{sec:CaseStudies}
\begin{figure}[t]
  \centering
  \begin{tabular}{@{}ccc@{}}
    \raisebox{1mm}{} &
    \includegraphics[width=80mm]{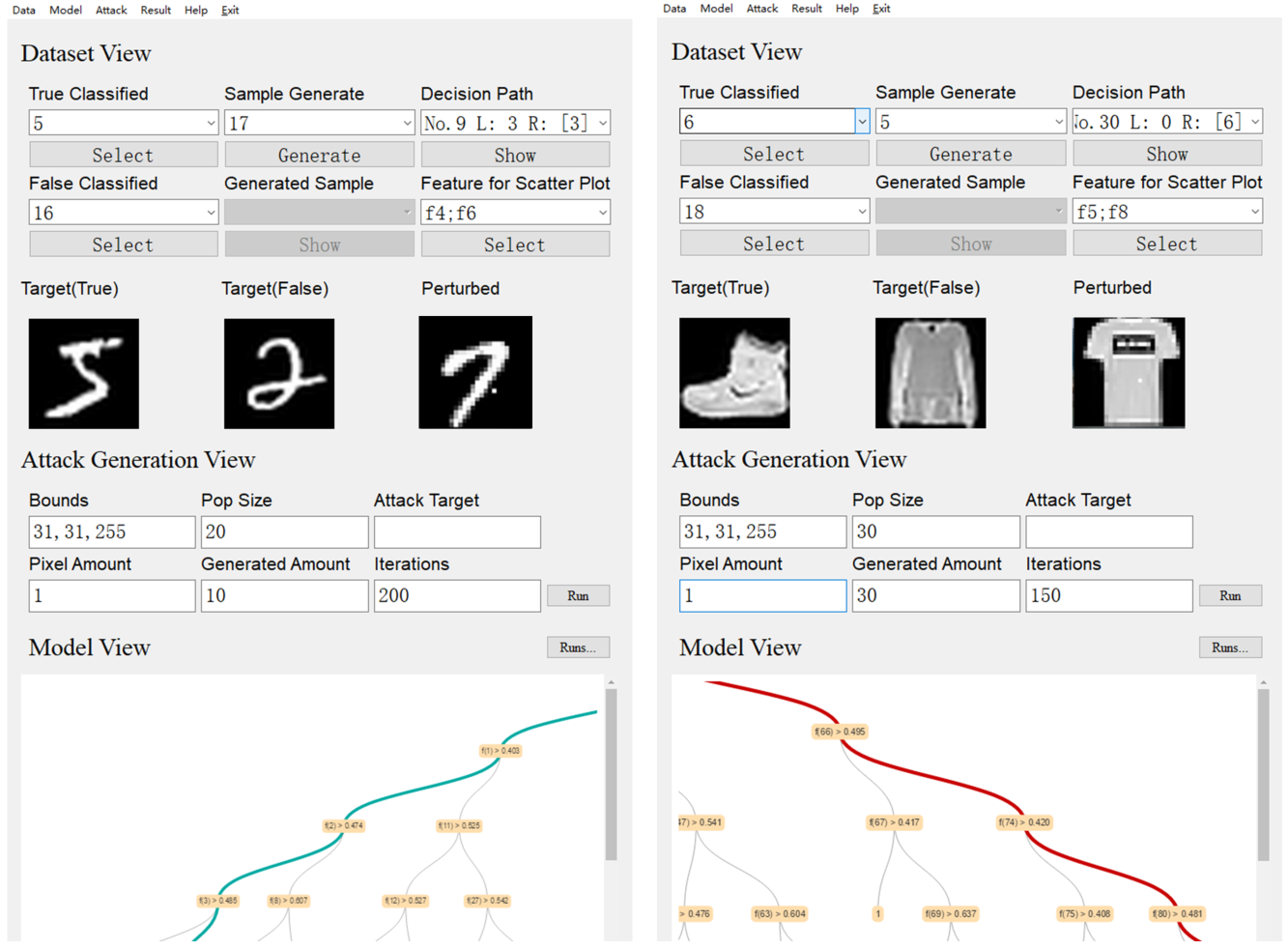} &
    \raisebox{1mm}{}
  \end{tabular}
  \caption{In addition to the Cifar-10 dataset featured in Fig. \ref{fig:MainScreen}, we also used left (a) the MNIST dataset and right (b) the Fashion-MNIST dataset for training ML models to be evaluated by TA3.}
  \label{fig:TwoDatasets}
\end{figure}

We evaluated TA3 software by testing classification models trained using three open ML datasets, i.e., the Cifar-10 dataset for 6 classes of animal images and 4 classes of transport images \cite{krizhevsky:2009:web} as shown in Fig. \ref{fig:MainScreen}, the MNIST dataset for 10 classes of handwritten digits \cite{Lecun:1998:IEEE} as shown in Fig. \ref{fig:TwoDatasets}(a), and the Fashion-MNIST dataset for 10 classes of clothing \cite{xiao:2017:Arx} as shown in Fig. \ref{fig:TwoDatasets}(b).

While the Cifar-10 dataset consists of color images, the MNIST dataset and Fashion-MNIST dataset consist of gray-scale images. Hence TA3 includes a modified version of the One Pixel attack algorithm, which perturbs gray-scale pixels instead of RGB pixels. This demonstrates the flexibility of TA3 in testing ML models trained with images in different color spaces.

Among these datasets, we trained several models to test various hypotheses about whether hypo-parameters used for the training processes would affect the models' resistance to adversarial attacks. When TA3 is used to test these models, it automatically generates a variety of plots as described in Section \ref{sec:Visualization}.   

\begin{figure}[t]
\centering
\begin{tabular}{@{}c@{\hspace{4mm}}c@{}}
    \small \textbf{attack breach rate} & \small \textbf{adversarial impact rate} \\
    \includegraphics[width=40mm]{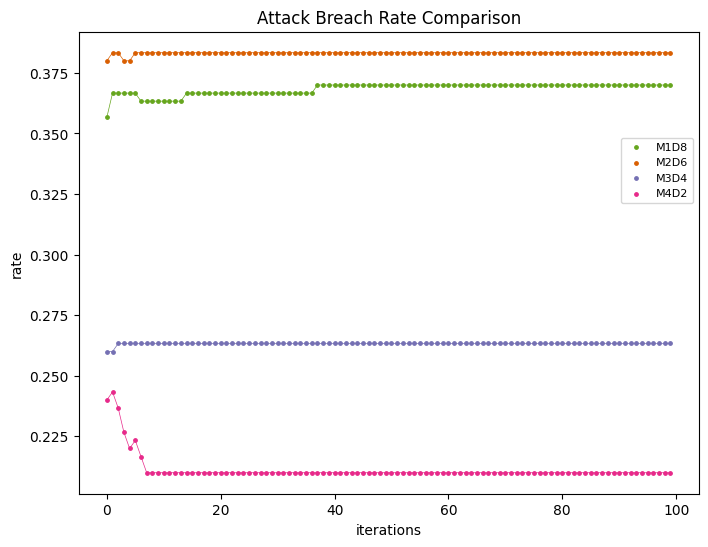} &
    \includegraphics[width=40mm]{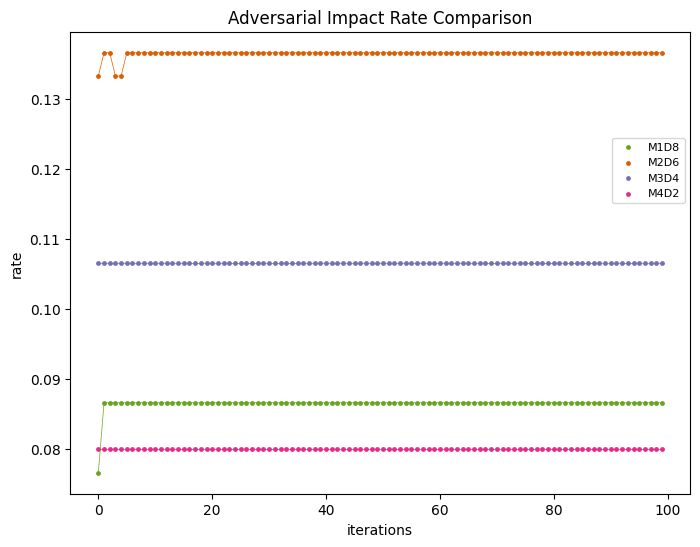} \\
    \small (a$_1$) Cifar-10 & \small (b$_1$) Cifar-10 \\[2mm]
    \includegraphics[width=40mm]{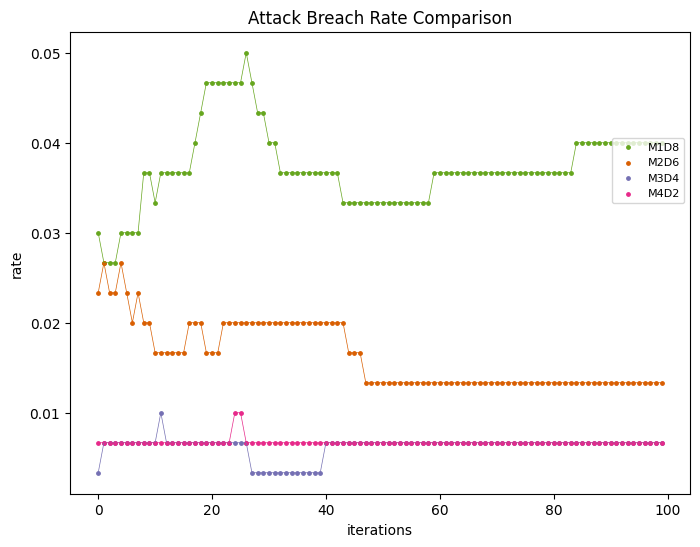} &
    \includegraphics[width=40mm]{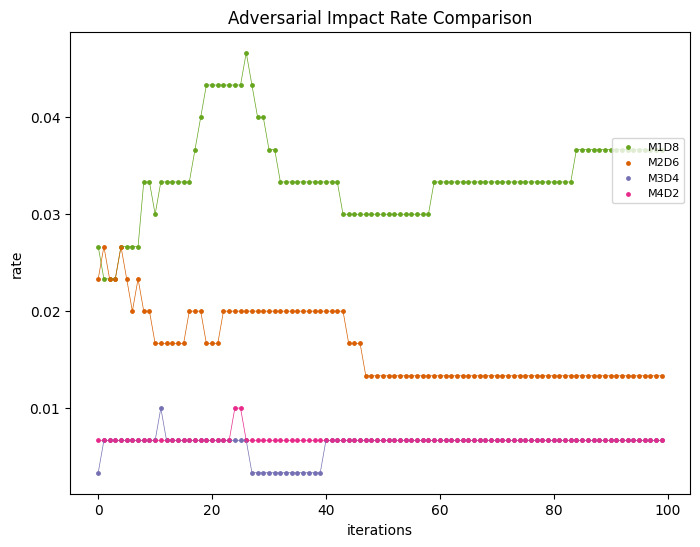} \\
    \small (a$_2$) MNIST & \small (b$_2$) MNIST \\[2mm]
    \includegraphics[width=40mm]{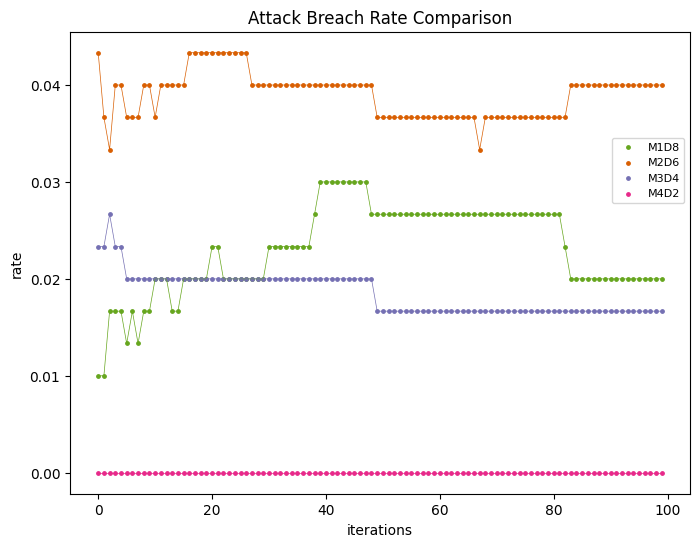} &
    \includegraphics[width=40mm]{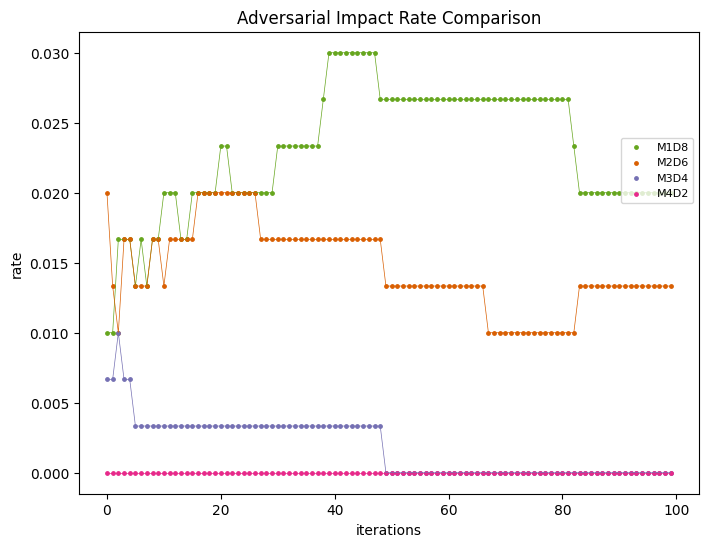} \\
   \small (a$_3$) Fashion-MNIST & \small (b$_3$) Fashion-MNIST
         %
    \end{tabular}
    \caption{Testing hypothesis 1 about depth control in training. Four models are of different tree depths, i.e., depth = 2 (M4D2, dark pink), 4 (M3D4, purple), 6 (M2D6, orange), and 8 (M1D8, green) respectively.
    (a) Comparing the \textbf{attack breach rates} of four models trained using each of the three datasets. (b) Comparing the \textbf{adversarial impact rates} of four models trained using each of the three datasets.
    }
    \label{fig:Plots}
\end{figure}

\textbf{Hypothesis 1: Does Depth Control Affect the Resistance?} For each of the three datasets, we trained four models that are of different tree depths, i.e., depth = 2, 4, 6, and 8 respectively. Fig. \ref{fig:Plots} shows two statistical measures, \textbf{attack breach rates} and \textbf{adversarial impact rates}, when these $4 \times 3$ models are tested iteratively, allowing us to observe and compare the performance of different models easily. 

It is common knowledge that the depth of a decision tree affects its accuracy. When a tree is too shallow, the model may suffer from under-fitting, and when it is too deep, it may be over-fitting. Consider 12 models, four trained with the Cifar-10 dataset (cM1D8, cM2D6, cM3D4, cM4D2), four trained with the MNIST dataset (mM1D8, mM2D6, mM3D4, mM4D2), and four trained with the Fashion-MNIST dataset (fM1D8, fM2D6, fM3D4, fM4D2). Here D2, D4, D6, and D8 stand for tree depths of 2, 4, 6, and 8 respectively. Note that we can train models with other depth control parameters, such as at depth 7, 10, 11, and so on, or dynamic depth control. In terms of accuracy, these models are ordered as:
%
%
\begin{align*}
\text{Cifar-10:} \quad      & Acc(cM2D6) > Acc(cM1D8) >\\
                            & Acc(cM4D2) > Acc(cM3D4)\\
\text{MNIST:} \quad         & Acc(mM1D8) > Acc(mM2D6) >\\
                            & Acc(mM3D4) > Acc(mM4D1)\\
\text{Fashion-MNIST:} \quad & Acc(fM1D8) > Acc(fM2D6) >\\
                            & Acc(fM3D4) > Acc(fM4D1)
\end{align*}

In this context, it would be interesting to find out whether depth control affects the model performance in terms of defending against adversarial attacks. From the plots in Fig. \ref{fig:Plots}, we can easily observe that models of different depths have different qualities in both \textbf{attack breach rates} and \textbf{adversarial impact rates}. However, their ordering may not always correspond to the accuracy ordering.

From Fig. \ref{fig:Plots}(a), we can observe that the models trained with the Cifar-10 and MNIST datasets apparently exhibit the same ordering in terms of \textbf{attack breach rates} and accuracy. In other words, the more accurate the model is, the easier it seems for attacks to be successful. Meanwhile, the \textbf{attack breach rates} for the models trained with the Fashion-MNIST dataset have a slightly different ordering from that of accuracy. This indicates that the two measures are not closely dependent.

As described in Section \ref{sec:Statistics}, an adversarial attack can sometimes ``correct'' an error made by a model (i.e., in the case of PNP and NPN). The measure \textbf{adversarial impact rates} excludes such events. 
From Fig. \ref{fig:Plots}(b), we can observe that the ordering of the four models changes for the models trained with the Cifar-10 and Fashion-MNIST datasets. In particular, comparing Fig. \ref{fig:Plots}(b$_1$) with Fig. \ref{fig:Plots}(a$_1$), the green line for M1D8 drops noticeable, indicating some successful attacks shown in a$_1$ are actually ``unintentional corrections'' of model errors. Meanwhile, comparing Fig. \ref{fig:Plots}(b$_3$) with Fig. \ref{fig:Plots}(a$_3$), we can observe that the orange line for M2D6 exhibits a similar phenomenon.
We can also observe that Fig. \ref{fig:Plots}(a$_2$) and Fig. \ref{fig:Plots}(b$_2$) are identical, indicating that no ``unintentional correction'' was found in testing models trained with the MNIST dataset.



\textbf{Hypothesis 2: Does Minimum Sample Split Affect the Resistance?} Minimum sample split is a parameter for controlling the minimum number of samples required to split an internal node \cite{Pedregosa:2011:ML}. We trained several models with the three datasets. Here we focused on four models (fM1S2, fM2S5, fM3S10, fM4S20) that were trained with the fashion-MNIST dataset, with the parameter for ``minimum sample split'' set as 2, 5, 10, and 20 respectively. The accuracy of the four models are (in fashion-MNIST):
\begin{align*}
    Acc(\text{fM1S2}) = 0.693 &> Acc(\text{fM2S5}) = 0.674 >\\
    Acc(\text{fM3S10}) = 0.659 &> Acc(\text{fM4S20}) = 0.642
\end{align*}

\begin{figure}[t]
    \centering
    \begin{tabular}{@{}c@{\hspace{1mm}}c@{}}
    \small \textbf{(a) Minimum Sample Split} & \small \textbf{(b) Maximum Features Selected} \\
    \includegraphics[width=40mm]{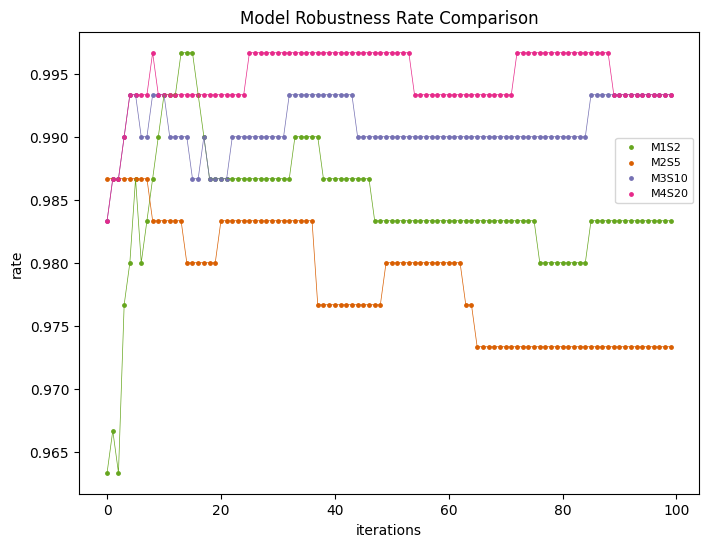} &
    \includegraphics[width=40mm]{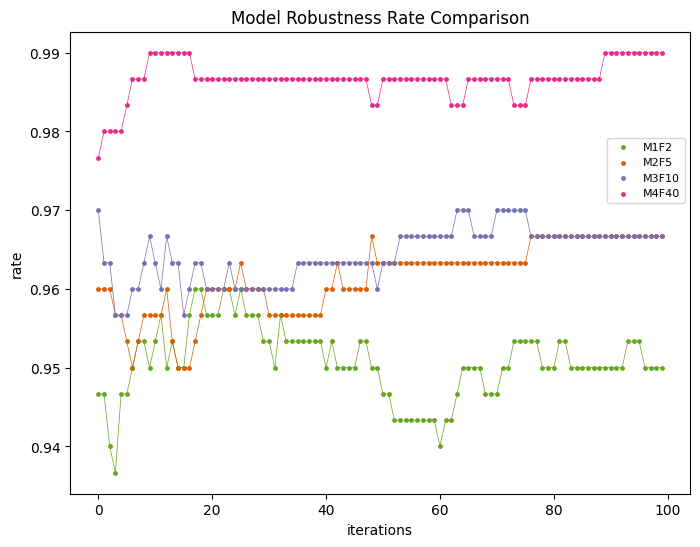} \\
    \multicolumn{2}{c}{\small (a$_1$) \hspace{4mm} Model robustness rate (MRR) \hspace{4mm} (b$_1$)}\\[2mm]
    \includegraphics[width=40mm]{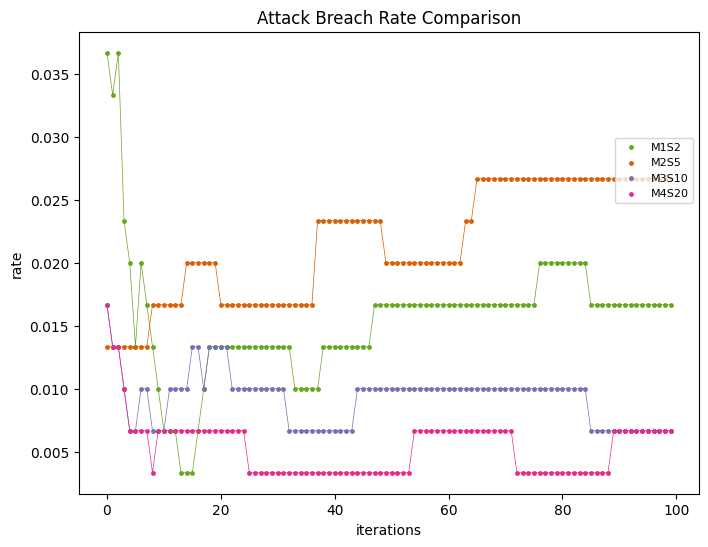} &
    \includegraphics[width=40mm]{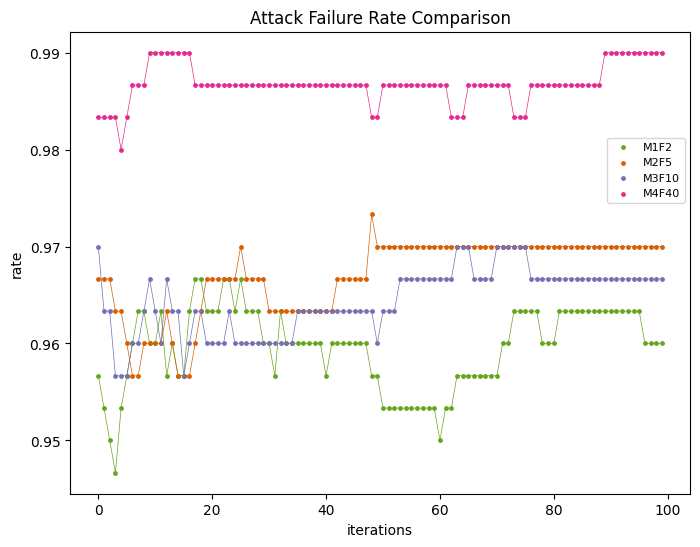} \\
    \multicolumn{2}{c}{\small (a$_2$) \hspace{7mm} Attack breach rate (ABR) \hspace{7mm} (b$_2$)}\\[2mm]
    \includegraphics[width=40mm]{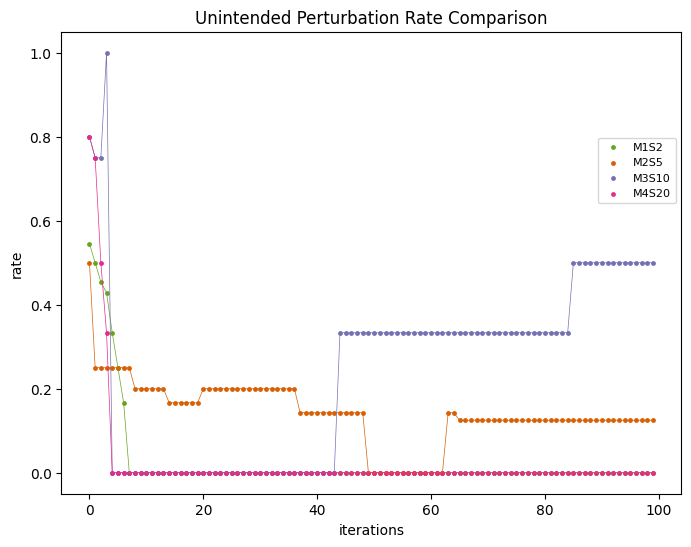} &
    \includegraphics[width=40mm]{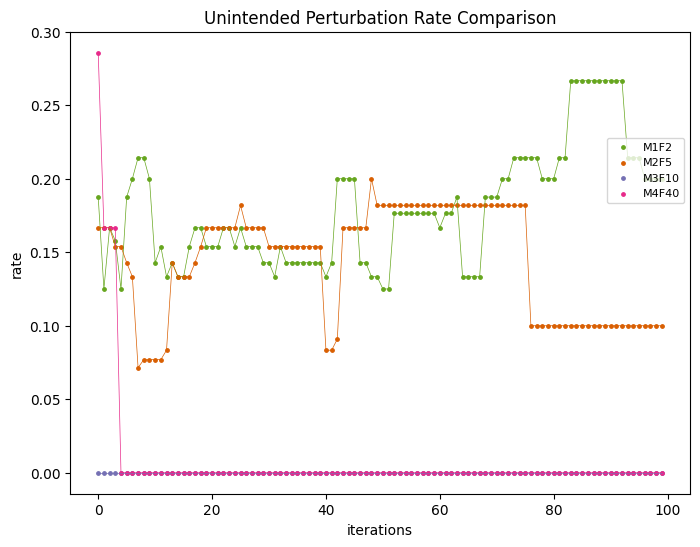} \\
    \multicolumn{2}{c}{\small (a$_3$) \hspace{1mm} Unintended perturbation rate (UPR) \hspace{1mm} (b$_3$)}
    \end{tabular}
    \caption{Testing hypotheses 2 and 3: (a) about minimum sample split, and (b) about maximum features selected. All models are trained with the fashion-MNIST dataset. In (a), four models trained with the parameter of minimum sample split is set to 2, 5, 10, and 20. In (b), Four models trained with the parameter of maximum features selected is set to 2, 5, 10, and 40.}
    \label{fig:TestH2}
\end{figure}

Fig. \ref{fig:TestH2}(a) provides evidence to support the hypothesis that changing this parameter can affect the models' ability against adversarial attacks. The three statistical indicators show patterns that are consistent with the order of their accuracy values. As shown in Fig. \ref{fig:TestH2}(a$_3$), while attacks on models fM1S2, fM2S5, and fM4S20 do not result much unintended perturbation, fM3S10 exhibits a rather different pattern at the later steps of the iteration. One possible explanation is that these ``positive attacks'' may happen in a certain area of the testing image, where fM3S10 tends to make errors.    

\textbf{Hypothesis 3: Does Maximum Features Affect the Resistance?} 
The maximum number of features considered by a training process is known to impact the quality of decision tree models \cite{Pedregosa:2011:ML}. We also trained several models with the three datasets. Here we focused on four models (fM1F2, fM2F5, fM3F10, fM4F40) that trained with the fashion-MNIST dataset, with the parameter for ``maximum features selected'' set as 2, 5, 10, and 40 respectively. The accuracy of the four models are (in fashion-MNIST):
\begin{align*}
    Acc(\text{fM1F2}) = 0.539 &< Acc(\text{fM2F5}) = 0.604 <\\
    Acc(\text{fM3F10}) = 0.625 &< Acc(\text{fM4F40}) = 0.693
\end{align*}

Fig. \ref{fig:TestH2}(b) provides evidence to support the hypothesis that changing this parameter can affect the models' ability against adversarial attacks. However, the three statistical indicators show patterns that are inconsistent with the order of their accuracy values. The \textbf{model robustness rate} (b$_1$) indicates \emph{MRR}(fM1F2) $<$ \emph{MRR}(fM2F5) $<$ \emph{MRR}(fM3F10) $<$ \emph{MRR}(fM4F40). In order words, this is consistent with the accuracy ordering. 
Meanwhile, the \textbf{attack failure rate} (b$_2$) shows that the positions of \emph{AFR}(fM2F5) and \emph{AFR}(fM3F10) are swapped.
On the other hand, the statistical measure of \textbf{unintended perturbation rate} (b$_3$) indicates that a good amount of attacks on fM1F2 and fM2F5 had ``positive'' outcomes.

\begin{figure}[ht]
    \centering
    \begin{tabular}{@{}c@{\hspace{2mm}}c@{\hspace{2mm}}c@{}}
         \includegraphics[width=28mm]{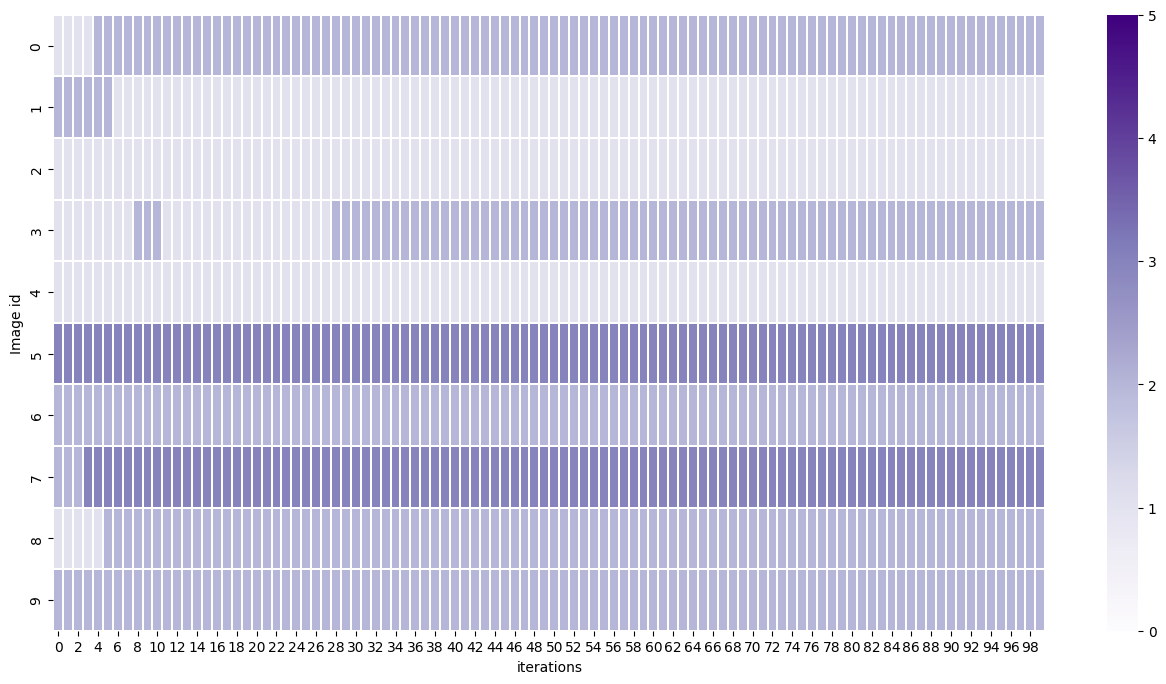} & 
         \includegraphics[width=28mm]{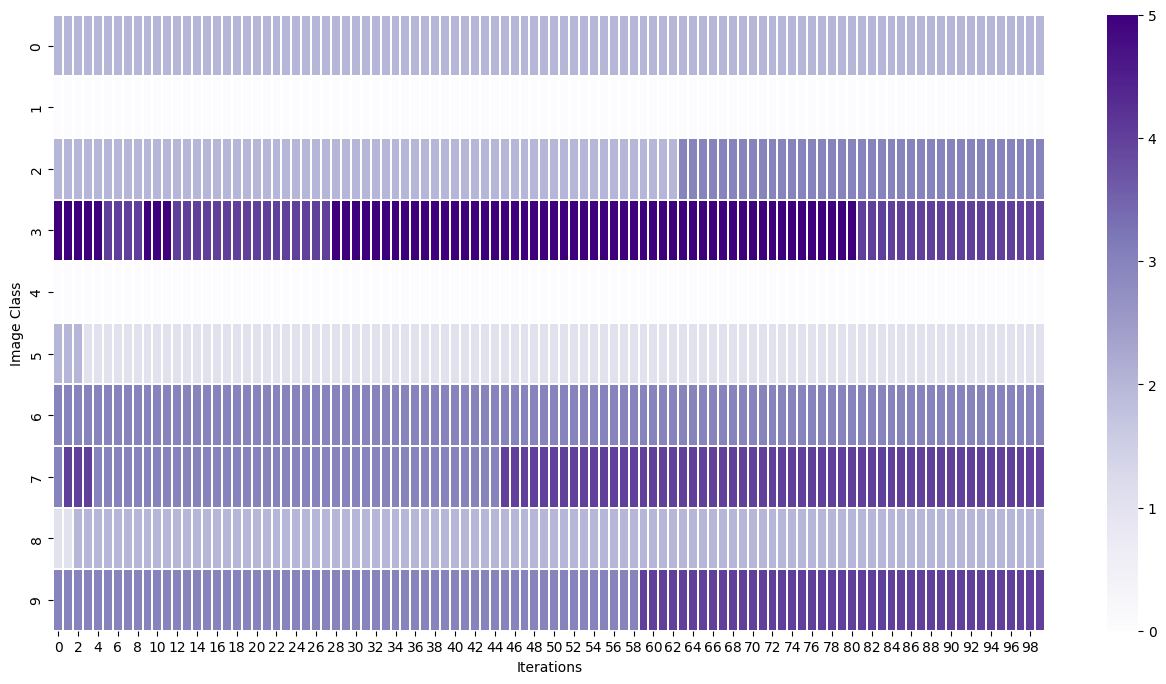} & 
         \includegraphics[width=28mm]{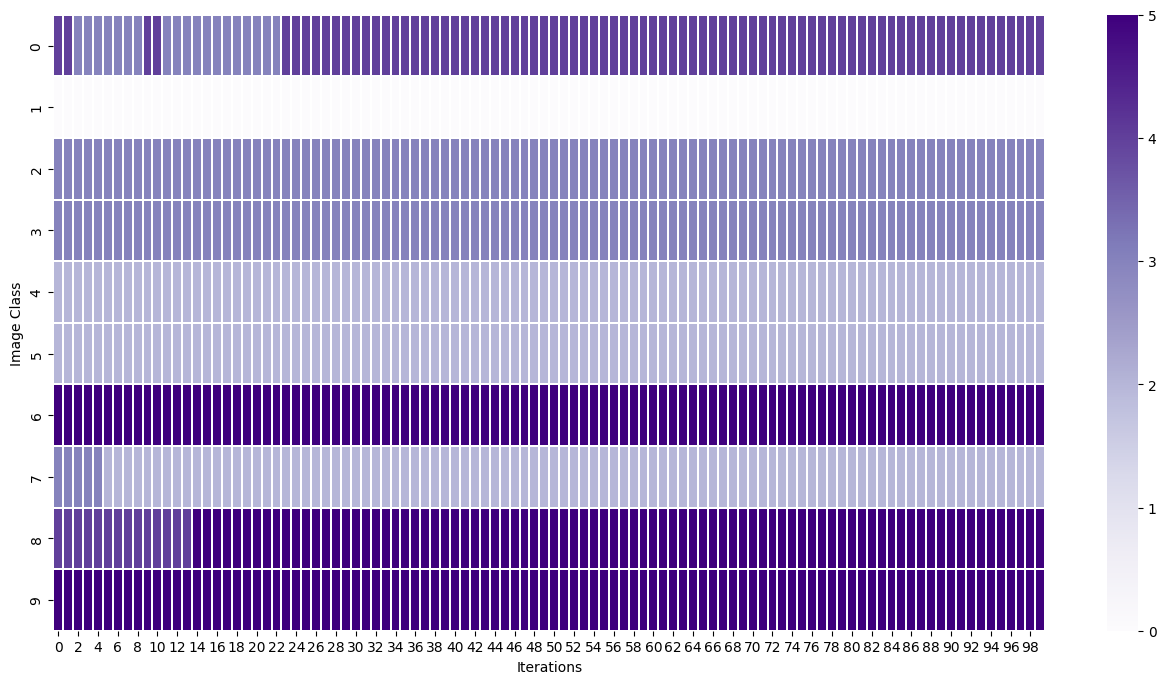} \\
         \small{(a$_1$) Cifar-10} &
         \small{(a$_2$) MNIST} &
         \small{(a$_3$) Fashion-MNIST}
    \end{tabular}
    \caption{Testing hypothesis 4 about image classes. In each plot, a row corresponds to a class, and the sequence of color-coded shapes in the row represents the varying number of successful attacks during the multi-iteration testing process.}
    \label{fig:TestH4}
\end{figure}

\textbf{Hypothesis 4: Do Image Classes Have Difference Vulnerability?}
Given a classification model, we can compare the vulnerability of different classes. Fig. \ref{fig:TestH4} shows the results of testing three models trained with the Cifar-10, MNIST, and Fashion-MNIST datasets respectively, The 10 classes of Cifar-10 are 0: airplane, 1: automobile, 2: bird, 3: cat, 4: deer, 5: dog, 6: frog, 7: horse, 8: ship, 9: truck. The 10 classes of MNIST are ten digits, i.e., 1, 2, 3, 4, 5, 6, 7, 8, 9, 0. The 10 classes of fashion-MNIST are 0: T-shirt, 1: trouser, 2: pullover, 3: dress, 4: coat, 5: sandal, 6: shirt, 7: sneaker, 8: bag, 9: ankle boot. In each plot in Fig. \ref{fig:TestH4}, the $k^{th}$ row shows the number of successful attacks to the images in the $k^{th}$ class when the testing progresses iteratively.

For each class, 100 images were randomly selected to be attacked. The numbers of successful attacks are color-coded, i.e., the darker the color is, the more attacks were successful and therefore the class concerned is potentially more vulnerable.
Fig. \ref{fig:TestH4} shows the results of testing each class in terms of the number of images (out of 100) that were breached in each iteration during the testing.
From Fig. \ref{fig:TestH4}, we can observe that color variation is more noticeable in (b) and (c), indicating that there is evidence for supporting the hypothesis. 



%% file: 7_conclusion.tex
\section{Discussions and Conclusion}
\textbf{Discussions.}
We recognize that testing against adversarial attacks is not an easy undertaking.
This work shows that there is a non-trivial amount and a variety of Human-in-the-loop (HITL) activities needed for supporting the testing against One Pixel attacks on decision tree models. This justifies the need for a purposely built user interface.

While some HITL activities are common to different ML models being tested or different adversarial attack algorithms, many will be specific to individual ML techniques and attack algorithms.
The strategy for developing a testing software system thus requires further research, to answer questions such as whether one system should accommodate different ML techniques and attack algorithms. Based on our experience in designing TA3, we suggest that for white-box testing, the best strategy might be to design multiple systems, one for each family of ML techniques that share the same form of internal structures. Meanwhile, such a system should include as many attack algorithms as possible that could be applied to the family of ML techniques. Meanwhile, for black-box testing, one may have a more generic user interface that includes many attack algorithms.

Technically, much more research and development effort will be needed to develop interactive software systems for different families of ML models, some of which will be challenging to visualize and interact with. There will also be a need for designing and developing an API to improve the efficiency of developing such interactive software systems. 

\textbf{Conclusions.}
In this paper, we present TA3, a novel HITL-support tool for testing ML models in terms of their ability against adversarial attacks. As illustrated in Fig. \ref{fig:Workflows}, the testing workflow (Fig. \ref{fig:Workflows}(b)) is much more complex than the one for testing an adversarial attack algorithm (Fig. \ref{fig:Workflows}(a)). This confirms the necessity of providing more efficient and effective HITL support. The development of TA3 confirms the feasibility of this approach. The design of the TA3 enables the close integration of algorithms, interaction, visualization, and statistics.



\textbf{Future Work.}
We plan to research the HITL support for testing other families of ML models, such as neural networks, Bayesian networks, and hidden Markov models, which feature different internal structures and hence need different support for interaction and visualization.
